\documentclass{aa}  

\usepackage{graphicx}
\usepackage{txfonts}
\usepackage{gensymb}
\usepackage{appendix}

\usepackage{natbib,twoopt}
\usepackage[breaklinks=true]{hyperref} 
\bibpunct{(}{)}{;}{a}{}{,} 
\makeatletter
\newcommandtwoopt{\citeads}[3][][]{\href{http://adsabs.harvard.edu/abs/#3}%
{\def\hyper@linkstart##1##2{}%
\let\hyper@linkend\@empty\citealp[#1][#2]{#3}}}
\newcommandtwoopt{\citepads}[3][][]{\href{http://adsabs.harvard.edu/abs/#3}%
{\def\hyper@linkstart##1##2{}%
\let\hyper@linkend\@empty\citep[#1][#2]{#3}}}
\newcommandtwoopt{\citetads}[3][][]{\href{http://adsabs.harvard.edu/abs/#3}%
{\def\hyper@linkstart##1##2{}%
\let\hyper@linkend\@empty\citet[#1][#2]{#3}}}
\newcommandtwoopt{\citeyearads}[3][][]%
{\href{http://adsabs.harvard.edu/abs/#3}
{\def\hyper@linkstart##1##2{}%
\let\hyper@linkend\@empty\citeyear[#1][#2]{#3}}}
\makeatother

\usepackage{enumitem}
\newlist{legal}{enumerate}{10}
\setlist[legal]{label*=\arabic*.}
\def\gaia{\textit{Gaia}\xspace}
\def\gdr3{\textit{Gaia}~DR3\xspace}
\def\g{$G$\xspace}
\def\bp{$G_{\rm BP}$\xspace}
\def\rp{$G_{\rm RP}$\xspace}
\def\bprp{\mbox{$G_{\rm BP}-G_{\rm RP}$}\xspace}

\titlerunning{\gaia~DR3: solar-like variability} 

\begin{document} 

\title{\gaia Data Release 3. Rotational modulation and patterns of colour variations in solar-like variables. }

   \author {E. Distefano
          \inst{1}
          \and
          A. C. Lanzafame
          \inst{1,2}
          \and
          E. Brugaletta
          \inst{1}
          \and
          B. Holl
          \inst{3}
          \and
          A. F. Lanza
          \inst{1}
          \and S. Messina
          \inst{1}
          \and I. Pagano
          \inst{1}
          \and M. Audard
          \inst{3}
          \and G. Jevardat de Fombelle
          \inst{3}
          \and I. Lecoeur-Taibi 
          \inst{3}
          \and N. Mowlavi
          \inst{3}
          \and K. Nienartowicz
          \inst{3,5}
          \and L. Rimoldini
          \inst{3}
          \and D. W. Evans,
          \inst{4}
           \and M. Riello
        \inst{4}
        \and P. García-Lario
            \inst{6}
          \and P. Gavras
          \inst{6}
        \and L. Eyer
          \inst{3}
           }

   \institute{ INAF - Osservatorio Astrofisico di Catania\\
              Via S. Sofia, 78, 95123, Catania, Italy\\
              \email{elisa.distefano@inaf.it}
            \and University of Catania, Astrophysics Section, Dept. of Physics and Astronomy\\
          Via S. Sofia, 78, 95123, Catania, Italy
          \and Department of Astronomy, University of Geneva, \\ Chemin Pegasi 51, 1290 Versoix, Switzerland
          \and Institute of Astronomy, University of Cambridge,\\ Madingley Road, Cambridge CB3 0HA, United Kingdom
          \and Sednai Sàrl, Geneva, Switzerland
          \and RHEA for European Space Agency (ESA), Camino bajo del Castillo,\\ s/n,Urbanizacion Villafranca del Castillo, Villanueva de la Ca\~nada, 28692 Madrid, Spain
       }

   \date{Received ; accepted }

 
  \abstract
   { The Gaia third Data Release (GDR3) presents a catalogue of 474\,026 stars (detected by processing a sample of about 30 millions late type stars) with variability induced by magnetic activity. About 430\,000 of these stars are new discovered variables. For each star, the catalogue provides a list of about 70 parameters among which the most important are the stellar rotation period $P$, the photometric amplitude $A$ of the rotational modulation signal and the Pearson Correlation Coefficient $r_0$ between magnitude and colour variations.}
   {In the present paper we highlight some features of the \gaia photometric time-series used   to obtain the catalogue and we present the main  attributes of the catalog. }
{ The Specific Objects Study (SOS) pipeline, developed to characterise magnetically active stars with Gaia Data, has been described in the paper accompanying the Gaia second Data Release. Here we describe the changes made to the pipeline and a new method developed to analyse Gaia time-series and to reveal spurious signals induced by instrumental effects or by the peculiar nature of the investigated stellar source. Such a method is based on the measurement of the per-transit-corrected-excess-factor ($c*$) for each  time-series transit, where $c*$ is a parameter that permits to check the consistency between $G$, \bp and \rp fluxes in a given transit. }
{The period-amplitude diagram obtained with the DR3 data confirms the DR2 findings i.e. the existence of a family of Low-Amplitude-Fast-Rotators never seen by previous surveys.
   The GDR3 data permitted, for the first time, to analyse the patterns of
   magnitude-colour variations for thousands of magnetically active stars. The measured $r_0$ values are tightly correlated with the stars position in the period-amplitude diagram. 
 }
  {The relationship between the $P$, $A$ and $r_0$ parameters inferred for thousands of stars could be very useful to improve the understanding of stellar magnetic fields and to improve theoretical models, especially in the fast rotation regime. The method developed to reveal the spurious signals can be applied to each of the released Gaia photometric time-series and can be exploited  by anyone  interested in working directly with Gaia time-series.}


   \maketitle
%

\section{Introduction}

The expression solar-like variables is commonly used to designate a wide class of objects (including late type dwarfs,T-Tauri and stars in RS CVn binary systems) whose variability is induced by the presence and the evolution of magnetically active regions (here after MARs). 
MARs are complexes of dark spots and bright faculae located in regions  with an enhanced magnetic field and unevenly distributed over the stellar photosphere.
In solar-like stars, the evolution of MARs is responsible for different variability phenomena whose time-scales can range between few minutes, as in the case of flare events, to years, as in the case of the 11-yr cycle observed in the Sun \citep[see e.g.][]{2004AN....325..660M}. 
Flares are outbursts induced by the magnetic reconnection, i.e. a rearrangement of the magnetic field topology occurring in a short time-scale \citep[see .e.g.][for a theoretical view of this mechanism]{1999Ap&SS.264..129S,2010ARA&A..48..241B}.  A  typical flare event exhibits a rapid (few minutes) increase of the stellar flux followed by a slower decay-phase whose duration can range between 20 minutes and 6 hours  \citep[see e.g.][]{2011AJ....141...50W,2016ApJ...829...23D,2019MNRAS.489..437D}.
At the other end of the scale, periodic or quasi-periodic variations in stellar flux can be induced by variations in the total area and geometrical distribution of MARs. Usually, these variations take place in time-scales of several years and are referred to as activity cycles.
At an intermediate time-scale, the typical signature of solar-like stars' light curves is the so-called rotational modulation signal, i.e. a quasi-periodic flux variation induced by the stellar rotation that modulates the visibility of MARs over the stellar disk.  
The analysis of the rotational modulation signal is a powerful tool to study the physical properties of a given star and to constrain theoretical models on stellar evolution and magnetic fields. 
Firstly, estimating the period of the signal  is  equivalent to measuring the stellar rotation period $P$. Secondly the amplitude $A$ of the signal can be used to constrain  the hemispheric asymmetry of MARs distribution, the MARs filling factor and the contrast between MARs and photosphere, that are the main factors driving the rotational modulation variability of the stellar flux. Thirdly, the temporal variations in the signal amplitude can be used to infer the time-scales over which the stellar magnetic field changes topology. Finally the estimate of $P$ in a large sample of stars with known ages allows us to investigate how the stellar rotation period depends on the stellar age and  to probe and improve the gyrochronology theories developed to model such a dependency \citep{1972ApJ...171..565S,2003ApJ...586..464B,2015A&A...584A..30L}.\\
Beside the period $P$ and the amplitude $A$ of the rotational modulation signal, another important diagnostic that can give precious information on the stellar magnetic field, is given by the Pearson correlation coefficient $r_0$ between magnitude and colour variations. Indeed this index can provide information on the contrast in temperature between the MARs and the stellar photosphere and on their geometrical distribution \citep[see e.g.][]{2008A&A...480..495M, 2019ApJ...879..114I}. Gaia, performing a multiband survey, permitted to measure $r_0$ and to study its correlation with $P$ and $A$ for an unprecedented number of stars.

In the present paper we illustrate the results obtained by searching for rotational modulation signals
in the GDR3 photometric time-series.
The analysis of GDR3 data allowed us to detect rotational modulation in a sample of 474\,026 stars, among which about 430\,000 are, as far as we know, new discovered variables. The results of our analysis are stored in the  GDR3 {\tt vari\_rotational\_modulation} catalogue (hereafter {\tt gdr3\_rotmod}), where we reported, for each star, a list of 66 parameters characterizing  their rotation and magnetic activity. 
The full description of these parameters can be found in the GDR3 documentation \citep{DR3-documentation}. 
Here we focus on the $P$, $A$ and $r_0$ distributions and in their reciprocal relationships.

In the second section of the paper we describe the pipeline used to detect and characterize the stellar magnetic activity. 
In Sec. 3, we present the main features of the catalogue and in Sec. 4 the conclusions are drawn.

In the three appendixes we analyse how the technical features of \gaia, like the time-sampling and the processing of photometric data, affect the completeness of the   {\tt gdr3\_rotmod} catalog. These appendixes are recommended to readers that are interested to a deeper understanding of \gaia data and to work with the released photometric time-series. 
In Appendix \ref{sec:ppf}, we illustrate in detail a method developed  to detect spurious signals in the \gaia photometric time-series. In Appendix \ref{qualityassesment}, we assess the completeness and the contamination of  {\tt gdr3\_rotmod} catalog.
Finally, in Appendix \ref{dr2dr3}, we present a comparison between the results on rotational modulation variables obtained  in GDR2 (\gaia second Data Release) and  GDR3, respectively. This analysis shows that, though DR3 provides a richer sample of rotational modulation variables, about 60\% of the variables detected in GDR2 could not be recovered in GDR3. The lack of detection of rotational modulation for these sources in GDR3 is due to several issues related to changes in the variability pipeline and in the strategy adopted to calibrate the photometric data. However, this does not imply that the rotational modulation data published in GDR2 are invalid.

\section{The method}
The pipeline used to search for the rotational modulation signal in the GDR3 time-series is a new version of that used for GDR2 and described in \cite{2018A&A...616A..16L}. The main  differences between the two versions of the code are that the quality criteria used to select the final sample of good candidates have been tightened and that the amplitude of the rotational modulation signal is now computed in all the three \gaia photometric bands.  
The  data processing can be outlined in six main steps:
\begin{itemize}
\item{sources selection }
\item{time-series segmentation}
\item{data cleaning}
\item{analysis of brightness-colour correlation}
\item{search for the rotational modulation signal}
\item{estimate of the stellar rotation period}
\end{itemize}
The pipeline was developed inside the framework of the Coordination Unit 7 (CU7) of the Gaia DPAC (Data Processing and Analysis Consortium). The CU7 softwares have the objective to analyse the photometric time-series collected by Gaia in order to classify and characterize the variability phenomena exhibited by the targeted stars. The raw time-series collected by \gaia are affected by bad measurements related to instrumental or calibration issues like those described in \cite{2021A&A...649A...3R}. These bad measurements can prevent the detection or can affect the characterisation of variability phenomena.  In order to mitigate the effects of these measurements, the CU7 team designed a chain of operators to clean the raw time-series and to reject all the outliers that are due to instrumental effects. The final output of this chain is given by the \texttt{ExtremeErrorCleaningMagnitudeDependent} time-series collected in the \gaia pass-bands \g, \bp and \rp (see \cite{2021A&A...649A...3R} and reference therein for details on the \gaia photometric system).  All the CU7 packages use these cleaned time-series as input. These time-series span an interval of 34 months. A full description of the chain of operators can be found in \cite{2022arXiv220606416E}.

\subsection{Data selection}
\label{datasel}
\begin{figure}
\begin{center}
\includegraphics[width=80mm]{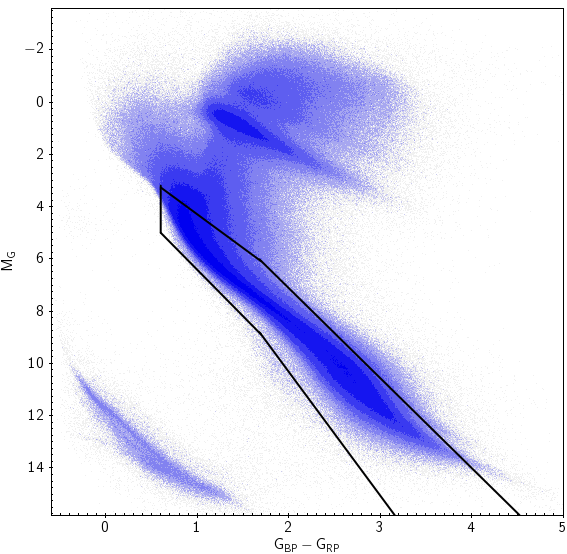}
\caption{{\tt DR3} $G$ vs.\bprp diagram. The black continuous lines mark the region used to perform a first selection of solar-like variables. }
\label{hr}
\end{center}
\end{figure}

The sources analysed in the present work are extracted from the {\tt geq5} catalogue described in \cite{DR3-DPACP-162}, i.e. the catalogue consisting of all the sources for which at least 5 non-null and positive FoV flux observations were available.
The {\tt geq5} catalogue lists about 1800 million sources. In order to save computing time, we run our pipeline only on a subset of about 30 million sources.
This subset was obtained by adopting the selection procedure described in \cite{2018A&A...616A..16L} and \cite{DR3-documentation}. 
\\
Briefly, a star is selected if it full-fills the following criteria:
\begin{enumerate}
\item{it has a positive parallax;}
\item{the relative error on parallax is less than 20\%; }
\item{its location in the $M_G$ vs. \bprp falls in the region delimited by the black-lines\footnote{the equations defining the region used for candidate selection  are provided in \cite{DR3-documentation}}  (Fig. \ref{hr}); }
\item{the sampling of its \g photometric time-series allows the extraction of at least one sub-series satisfying the condition
\begin{equation}
L \le 120~d \mbox{~and~} N_{P} \ge 12
\end{equation}
}
\end{enumerate}
where $L$ is the time interval spanned by the sub-series and $N_P$ is the number of photometric measurements inside the sub-series.
The first, second and third criteria are applied in order to limit the analysis only to stars whose location in the $M_G$ vs. \bprp is consistent with dwarfs having spectral type later than F5 or with T-Tauri objects. Obviously, this preliminary filtering is very rough and the region used for selection is expected to include several types of contaminants.
Most of these contaminants will be removed by the subsequent steps of the pipeline. The final percentage of contaminants in the released sample of variables is evaluated in Appendix \ref{contamination}. 
The fourth criterion is used to select the stars whose photometric time-series have a sampling suitable to the detection of the rotational modulation signal, as discussed in the next section.

\subsection{Segmentation}
\label{sec::segmentation}
The amplitude and the shape of the rotational modulation signal change in time because of the intrinsic evolution of MARs.
The time-scale over which the signal is stable and coherent is closely related to the typical life-time of MARs $\tau_{\rm MAR}$. 
Ideally, the search for the rotational modulation signal should be performed in sub-series whose length does not exceed $\tau_{\rm MAR}$.

The estimate of this time-scale has been the subject of several works and is related to different parameters as the stellar age and the spectral type.
The analysis of the TSI (Total Solar Irradiance) and SSI (Solar Spectral Irradiance) time-series showed that in the Sun the typical lifetime of spots $\tau_{\rm spot}$ is about 9 days whereas the typical lifetime of faculae $\tau_{\rm fac}$ is about 60 days, that roughly corresponds to two solar rotations \citep{2004A&A...425..707L}. 
In the same work, the authors show that, in some cases, the rotational modulation signal in the Sun can be coherent up to 150 d because spots and faculae tend to occur at preferential longitudes  giving rise to ensembles of MARs   whose typical lifetime $\tau_{\rm MAR}$ is about 200-250 d. However the authors point out that the rotational modulation signal of the Sun can be detected, through period search algorithms, only in 150~d intervals close to the minima of the 11-yr activity cycle. In fact in these intervals, the Sun activity is dominated by faculae and therefore the rotational modulation signal is less affected by the luminosity dips induced by spots.  These findings were recently confirmed by \cite{2017NatAs...1..612S}.      

The MARs lifetime for cool stars other than the Sun has been estimated for  limited sample of stars by \cite{1997SoPh..171..191D,1997SoPh..171..211D}, \cite{2002AN....323..349H} and \cite{2003A&A...409.1017M}.
\cite{1997SoPh..171..191D,1997SoPh..171..211D} found that in old stars, characterized by a low magnetic activity level, $\tau_{\rm MAR}$ tends to be shorter than the stellar rotation period preventing the detection of the rotational modulation signals, whereas
in young and magnetically active stars $\tau_{\rm MAR}$ values are of the order of several stellar rotations and range between 50 d and 1 year. Similar values are also reported by \cite{2002AN....323..349H} and by \cite{2003A&A...409.1017M}.
On the basis of these works, \cite{2016A&A...591A..43D} demonstrated that segmenting long-term photometric time-series in 50-d intervals can be a good strategy to correctly retrieve the rotational modulation signal in magnetically active stars. 
 Unfortunately, because of the \gaia time-series sampling, a 50 d interval could not have  enough transits\footnote{The authors remind that, in this work, the term transit  indicates the crossing of the entire focal plane by a given source. During a given transit \gaia collects a set of four measurements with the different instruments on board of the satellite \citep[see][for further details on \gaia instrumentation]{2016A&A...595A...2G}. }to perform a meaningful period search.   
A compromise between the sub-series length and the number of transits per-sub-series has therefore to be found. Moreover, the sampling changes with the star coordinates according to the \gaia scanning law \citep[see][for further details]{2005MNRAS.361.1136E,2012MNRAS.421.2774D}.    
In order to take into account the features of the \gaia sampling and the intrinsic evolution of MARs, the \g, \bp  and \rp photometric time-series of the selected target stars 
are divided in sub-series satisfying the  condition 4, reported in Sec .\ref{datasel}, through the  adaptive segmentation algorithm described in \cite{2018A&A...616A..16L}. This algorithm is adaptive because the segmentation strategy changes with the sky regions in which the analysed stars are located.

\begin{figure}
\begin{center}
\includegraphics[width=80mm]{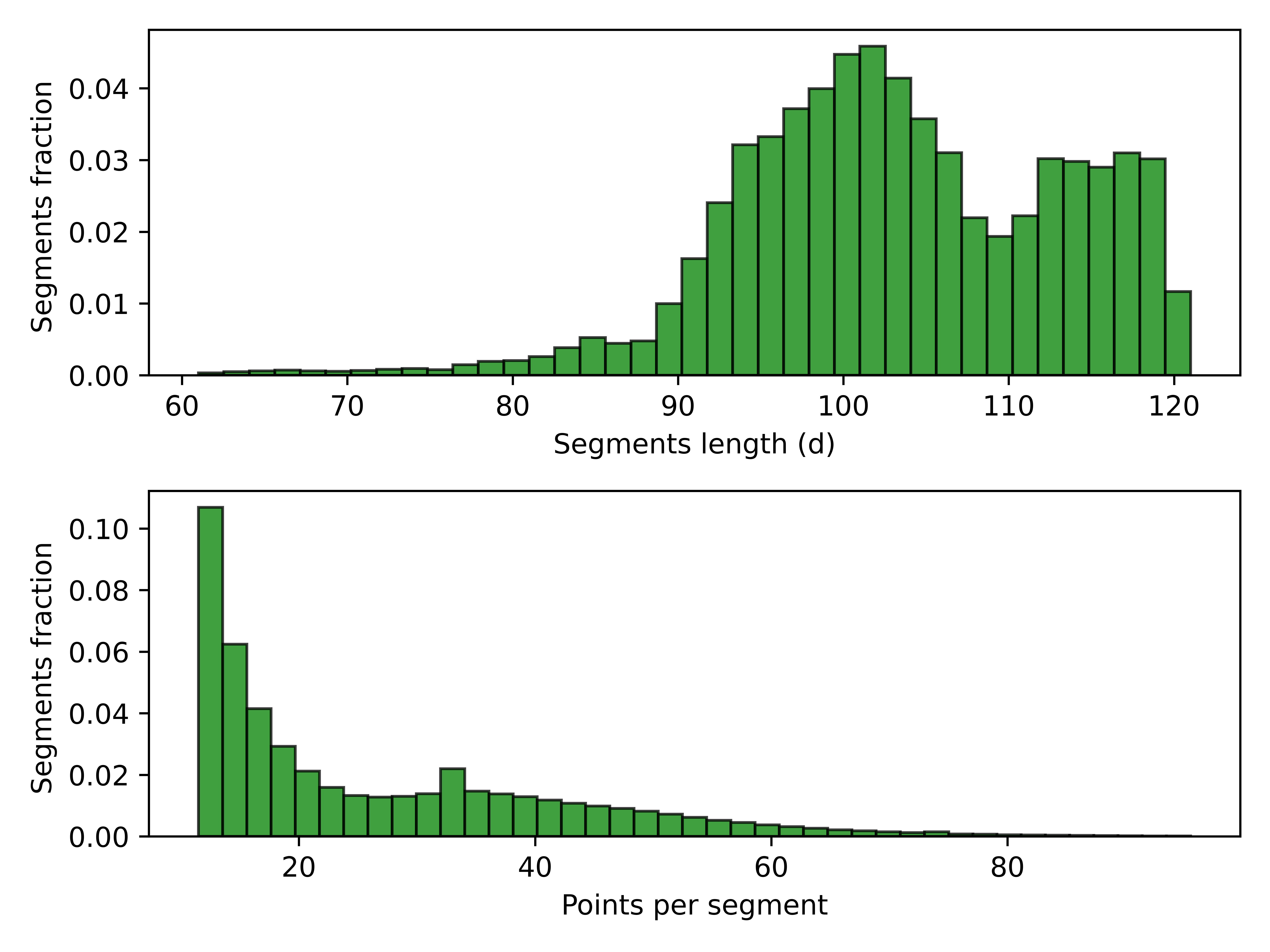}59
\caption{Top panel:  distribution of the segment length $L$. Most of the processed segments have a length $L=100~d$.
Bottom panel:  distribution of the number of points per segment $N_P$.}
\label{histoseg}
\end{center}
\end{figure}

In the top and bottom panel of Fig.\ref{histoseg} we report the distribution of the sub-series length and of the number of transits per sub-series, respectively. Most of the segments are characterized by a 100 d length and include 12 transits but there is also a small fraction of segments shorter than 20 d and a small fraction of segments with more than 100  points.

\begin{figure*}
\begin{center}
\includegraphics[width=160mm]{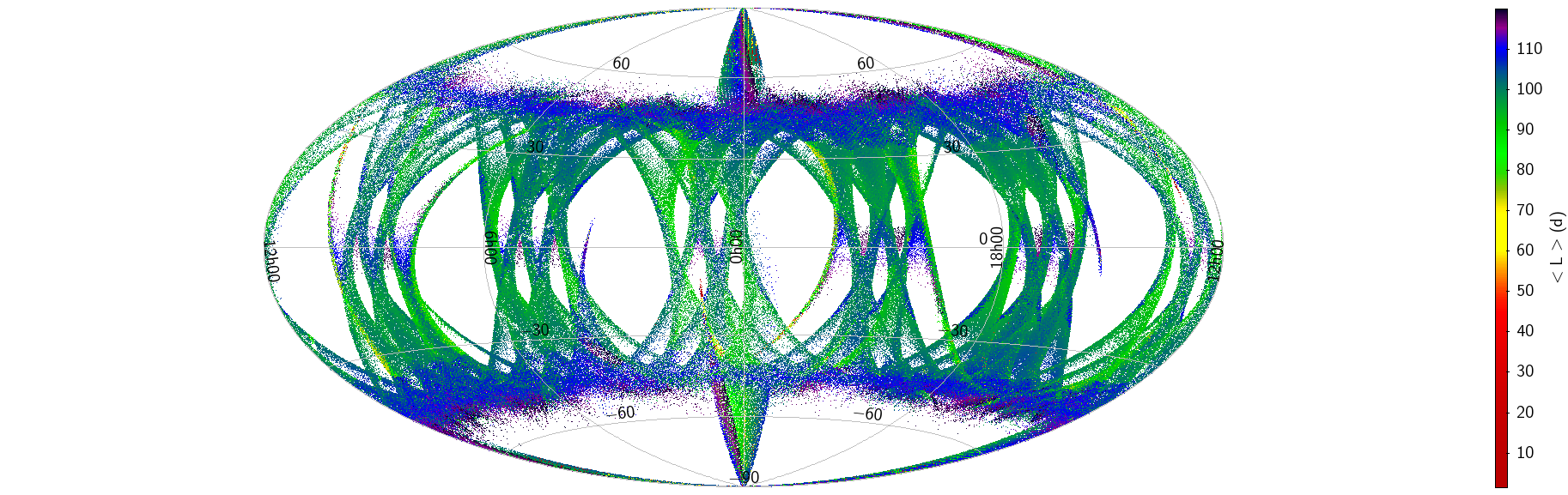}
\caption{Distribution map (in ecliptic coordinates) of the average segment length $<L>$.}
\label{mapavglen}
\end{center}
\end{figure*}

\begin{figure*}
\begin{center}
\includegraphics[width=160mm]{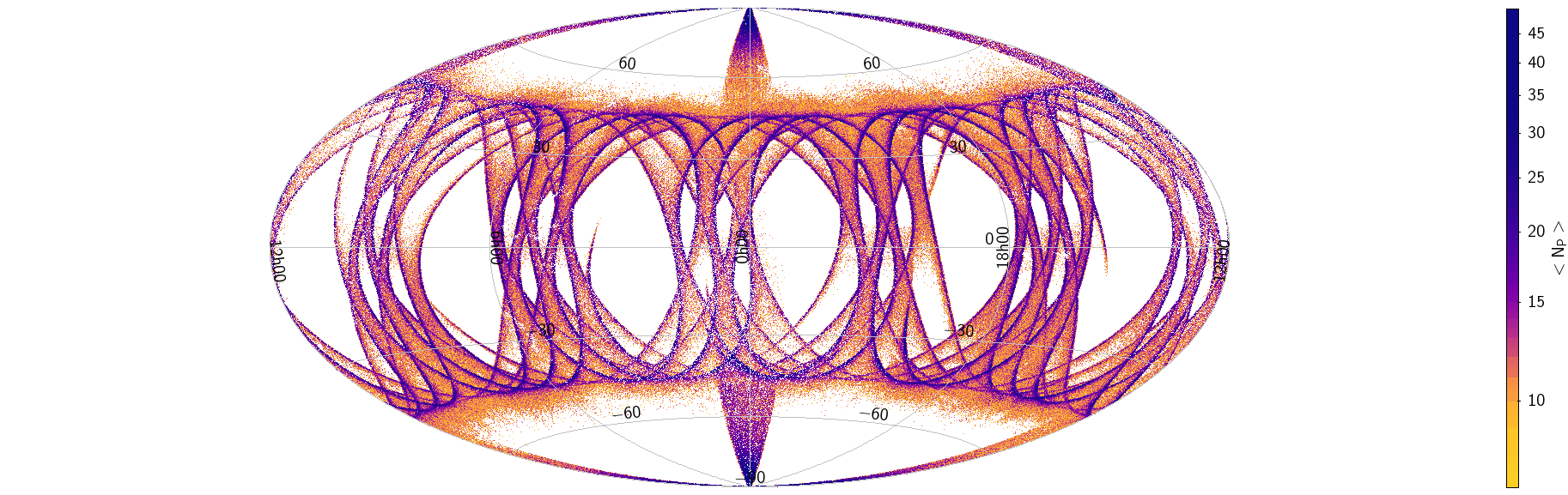}
\caption{Distribution map (in ecliptic coordinates) of the average number of points per segment $<N_P>$.}
\label{mapavgnp}
\end{center}
\end{figure*}

The distribution of $N_{\rm P}$ and $L$
changes with the sky location of the targets stars. In Fig. \ref{mapavglen} and \ref{mapavgnp} we report the median values of $N_{\rm P}$ and $L$ as a function of the ecliptic coordinates. The stars located at ecliptic latitudes close to $\beta\pm 45^{\circ}$ are scanned more frequently by the satellite and the corresponding sub-series have a higher number of transits $N_{\rm P}$.

\begin{figure*}
\begin{center}
\includegraphics[width=160mm]{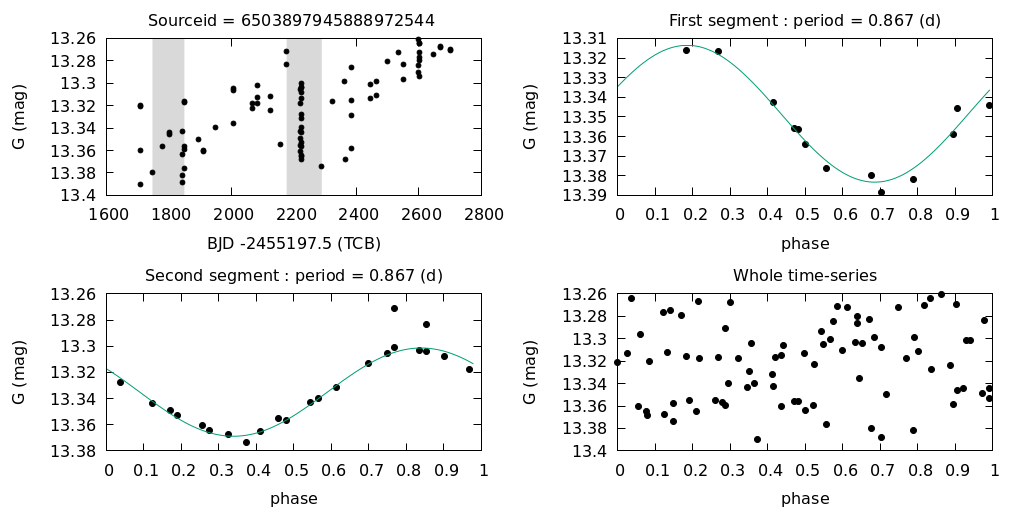}
\caption{Top left panel: full $G$ time-series for the star {\tt Gaia DR3 6503897945888972544}. The gray shadows mark two of the sub-series extracted by the segmentation algorithm. The period search algorithm retrieved a period $P=0.867 d$ in both segments. Top right panel: the first sub-series folded according to the period $P=0.867$. Bottom left panel: the second sub-series folded according to $P=0.867 d$. Bottom right panel: the full time-series folded according $P=0.867 d$. The rotational modulation signal looses coherence across the full time-series because of the intrinsic evolution of MARs and can be detected only in the shorter sub-series.  }
\label{example1}
\end{center}
\end{figure*}

\begin{figure*}
\begin{center}
\includegraphics[width=160mm]{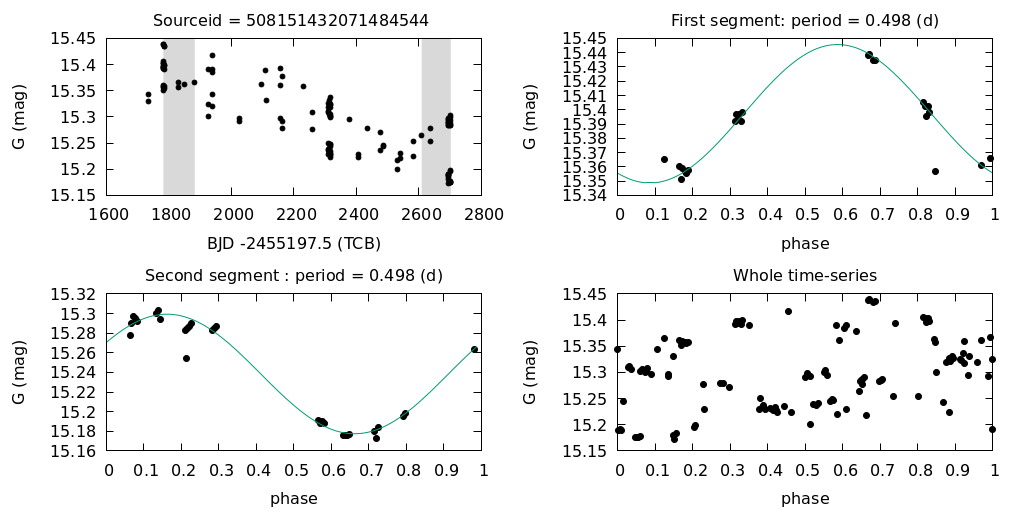}
\caption{Same of Fig. \ref{example1} for the star {\tt Gaia DR3 508151432071484544}. Also in this case the rotational modulation signal  looses coherence across the full time-series. }
\label{example2}
\end{center}
\end{figure*}

\begin{figure*}
\begin{center}
\includegraphics[width=160mm]{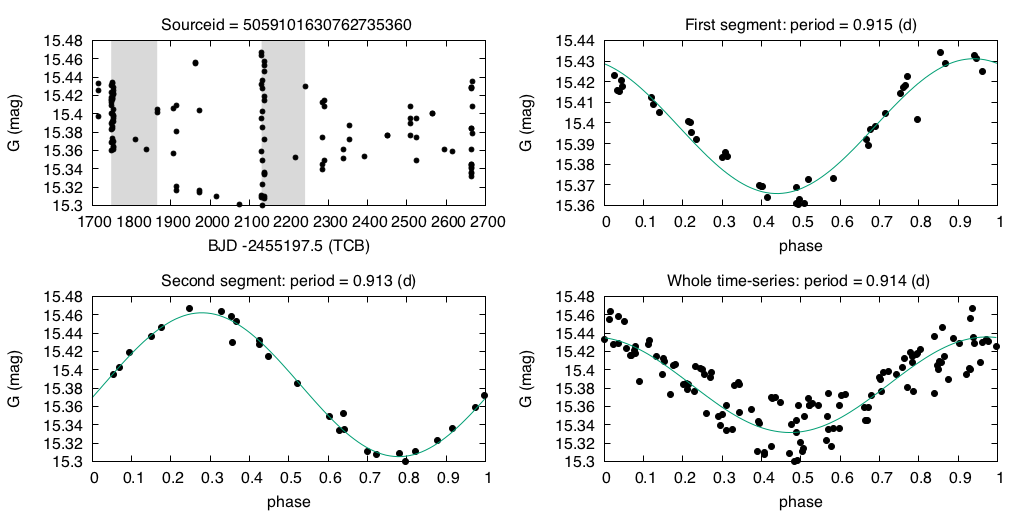}
\caption{Same of Fig. \ref{example1} for the stars {\tt Gaia DR3 5059101630762735360}. In this case the rotational modulation signal is coherent across the full time-series, indicating that the MARs are characterized by a long-term stable pattern.}
\label{example3}
\end{center}
\end{figure*}

\begin{figure*}
\begin{center}
\includegraphics[width=160mm]{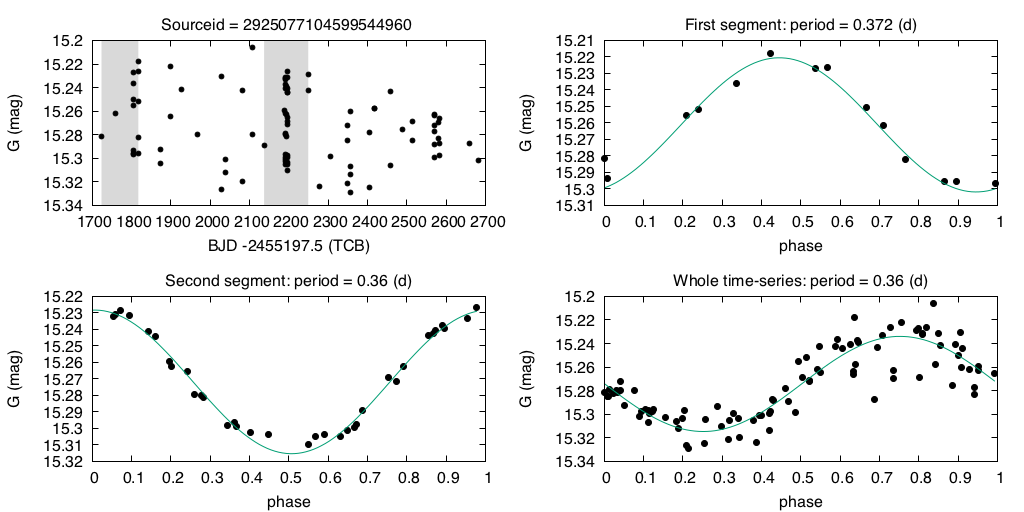}
\caption{Same of Fig. \ref{example1} for the star {\tt Gaia DR3 2925077104599544960}. In this case the rotational modulation signal is stable across the full time-series. }
\label{example4}
\end{center}
\end{figure*}

In Fig. \ref{example1}  we illustrate a typical case in which the time-series segmentation is crucial to retrieve the stellar rotation period. In the top-left panel of the plot, we display the whole \g time-series of the source {\tt Gaia  DR3 6503897945888972544}. The gray shaded regions indicate two of the sub-series generated by the segmentation algorithm (note that the other segments extracted by the pipeline are not marked in order to make the plot clearer). In both segments, a signal with period $P~=~0.867~d$ was detected. In the top-right  and bottom-left panels, we report the folded \g sub-series corresponding to the first and second segment, respectively. In the bottom-right panel we report the whole time-series folded according to the same period $P~=0.867~d$. While in the two sub-series the  modulation signal is coherent and stable, in the whole time-series it loses coherence because  MARs evolution induces a shift in phase between the two segments. Although the signal cannot be detected in the whole time-series, the segmentation algorithm allows us to retrieve the stellar rotation period. A similar case is reported in Fig. \ref{example2}. 
Note that MARs evolution not always destroys the signal coherence. In Fig. \ref{example3} we illustrate the case of the star {\tt Gaia DR3 5059101630762735360}. In this star, the amplitude of the signal increases in time and changes from 0.07 mag in the first segment to 0.13 mag in the second one. Nevertheless, the signal is still detectable in the whole time-series. 
Finally, there are stars in which the signal is stable and coherent along the whole time-series as in {\tt Gaia DR3 2925077104599544960} (see Fig. \ref{example4}).
At the end of this section we want to point out  that, recently, \cite{2022ApJ...924...31B} conducted an extensive study based on a sample of about 60\'000 stars for which both {\it Kepler} photometric time-series and \gaia stellar parameters were available. 
\cite{2022ApJ...924...31B} found that the $\tau_{\rm spot}$ distribution ranges between 10 and 350 d and it is peaked at around 140 d for young stars rotating faster than Sun, whereas it ranges between 10 and 250 d and is peaked at 80 d for stars older or as old as the Sun. Unfortunately, this work was not available at the time in which DR3 data were processed (i.e. December 2020). In the future \gaia releases, it could be very helpful to redefine the segmentation strategy.

\subsection{Data cleaning}
\label{sec:cleaning}
Once the time-series have been segmented, each sub-series is cleaned from spurious data. 
A first cleaning procedure is performed by rejecting all transits for which one or two of the $G$, \bp, \rp measurements are missing. This cleaning is performed in order to assure that the magnetic activity indexes computed in the different photometric bands are inferred from the same set of transits.
Once transits with missing measurements are removed, the pipeline searches for outliers due to possible flare events according to the procedure described in \cite{2018A&A...616A..16L} and in Sec.  10.14.3 of \cite{DR3-documentation}.
These outliers are flagged as candidate flare events and removed from the sub-series because they could prevent the detection of the rotational modulation signal.
\subsection{Analysis of the correlation between brightness and colour variations}
\label{sec:colorbrightness}
The cleaned time-series are analysed to investigate the correlation between the brightness and the colour variations in the selected sample of stars. 
For each  star and for each segment, the pipeline performs a robust linear regression  between the $G$ and \bprp measurements collected in the segment and estimates the slope $s$ and the intercept $i$ of the straight line best-fitting the data, \citep[see][for details on the algorithm employed to compute the regression]{2018A&A...616A..16L}.
For each segment, the Pearson Correlation Coefficient $r_0(G, (G_{\rm BP}-G_{\rm RP}))$ and its associated $p$ value are also computed. The closer $| r_0 |$ is to 1, the higher the strength of the linear correlation between the  colour and the magnitude variations. The  $p$ value gives the probability that the measured $r_0$ is obtained by chance. Hence, the lower  $p$ is, the higher  the statistical significance of the correlation.
For a given star, the $s$, $i$, $r_0$ and $p$ parameters are stored in the arrays:
\begin{itemize}
    \item {\tt segments\_color\_mag\_slope},
    \item {\tt segments\_color\_mag\_intercept},
    \item {\tt segments\_correlation\_coefficient},
    \item {\tt segments\_correlation\_significance}
\end{itemize}

of the {\tt gdr3\_rotmod}, respectively.
Note that these parameters were estimated  and reported also for the segments in which a meaningful rotational modulation signal is not detected.
The analysis of these parameters shows that there are roughly three distinct families of stars  that, according to the definition given by \cite{2008A&A...480..495M}, we call:
\begin{itemize}
    \item{ {\tt reddening-color-magnitude-correlated}  (hereafter RCMC) stars};
    \item{{\tt blueing-color-magnitude-correlated} (here after BCMC) stars};
    \item{{\tt color-magnitude-uncorrelated} (here after UMC stars)};
\end{itemize}
In  RCMC stars the colour and the magnitude variations are positively correlated, i.e.  as their brightness decreases their colour get redder.
In BCMC stars the colour and the magnitude variations are anti-correlated, i.e. as they get fainter they become bluer.
Finally in UMC stars the colour and magnitude variations are poorly correlated.
The transition between the three families is very smooth. 
In some RCMC stars the positive correlation is well defined in all the segments. In other RCMC stars a significant correlation is observed only in some segments. The same applies to BCMC stars.
Finally there are stars that in some segments are reddening and in others are blueing.

\begin{figure*}
\begin{center}
\includegraphics[width=160mm]{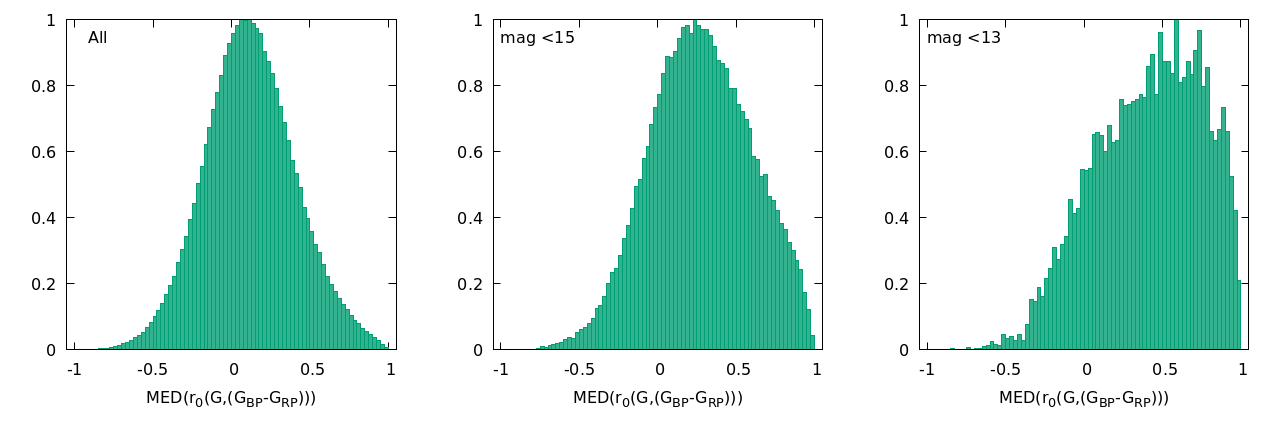}
\caption{normalised distribution of $MED(r_0(G, (G_{\rm BP)} -G_{\rm RP})))$ for all the stars of the {\tt gdr3\_rotmod} catalogue (left panel), for stars with $G < 15$ (central panel) and for stars with $G <13$ (right panel).}
\label{histor0}
\end{center}
\end{figure*}

In Fig. \ref{histor0} we present the normalised distribution of the median Pearson Correlation Coefficient $MED(r_0(G,(G_{\rm BP} -G_{\rm RP})))$ in different range of magnitudes: as the stars get fainter, the percentage of uncorrelated stars increases because the photometric noise could mask the correlation between colour and magnitude variations.  Note that these distributions are built without taking into account the statistical significance
associated with $MED(r_0(G,(G_{\rm BP} -G_{\rm RP})))$
otherwise the percentage of uncorrelated stars will be under-estimated.

\begin{figure}
\begin{center}
\includegraphics[width=80mm]{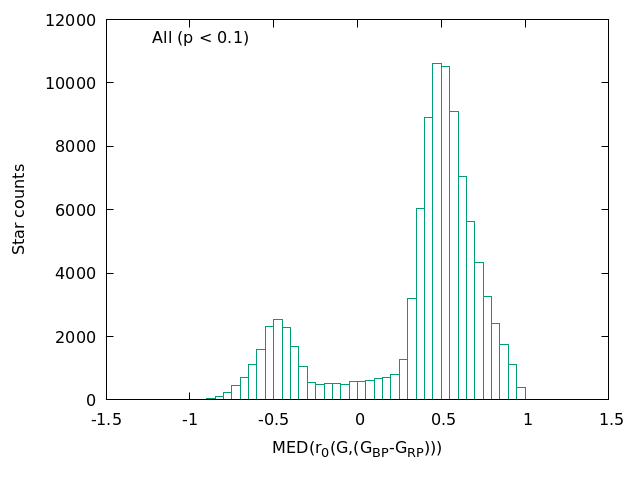}
\caption{Distribution of the median Pearson correlation coefficient $MED(r_0(G,(G_{\rm BP}-G_{\rm RP})))$ for all the targets in which $p < 0.1$. }
\label{histor0sig}
\end{center}
\end{figure}

In Fig. \ref{histor0sig} we report the distribution of $MED(r_0(G,(G_{\rm BP} -G_{\rm RP})))$ for all the stars in which its associated $p$ value is less than 0.1. The histogram shows that the blueing stars with an high statistical significance are only a small percentage ($\simeq~4\%$) of the entire {\tt gdr3\_rotmod} catalog, whereas the reddening stars are more numerous and are about the $17\%$ of the catalog.

\begin{figure*}
\begin{center}
\includegraphics[width=160mm]{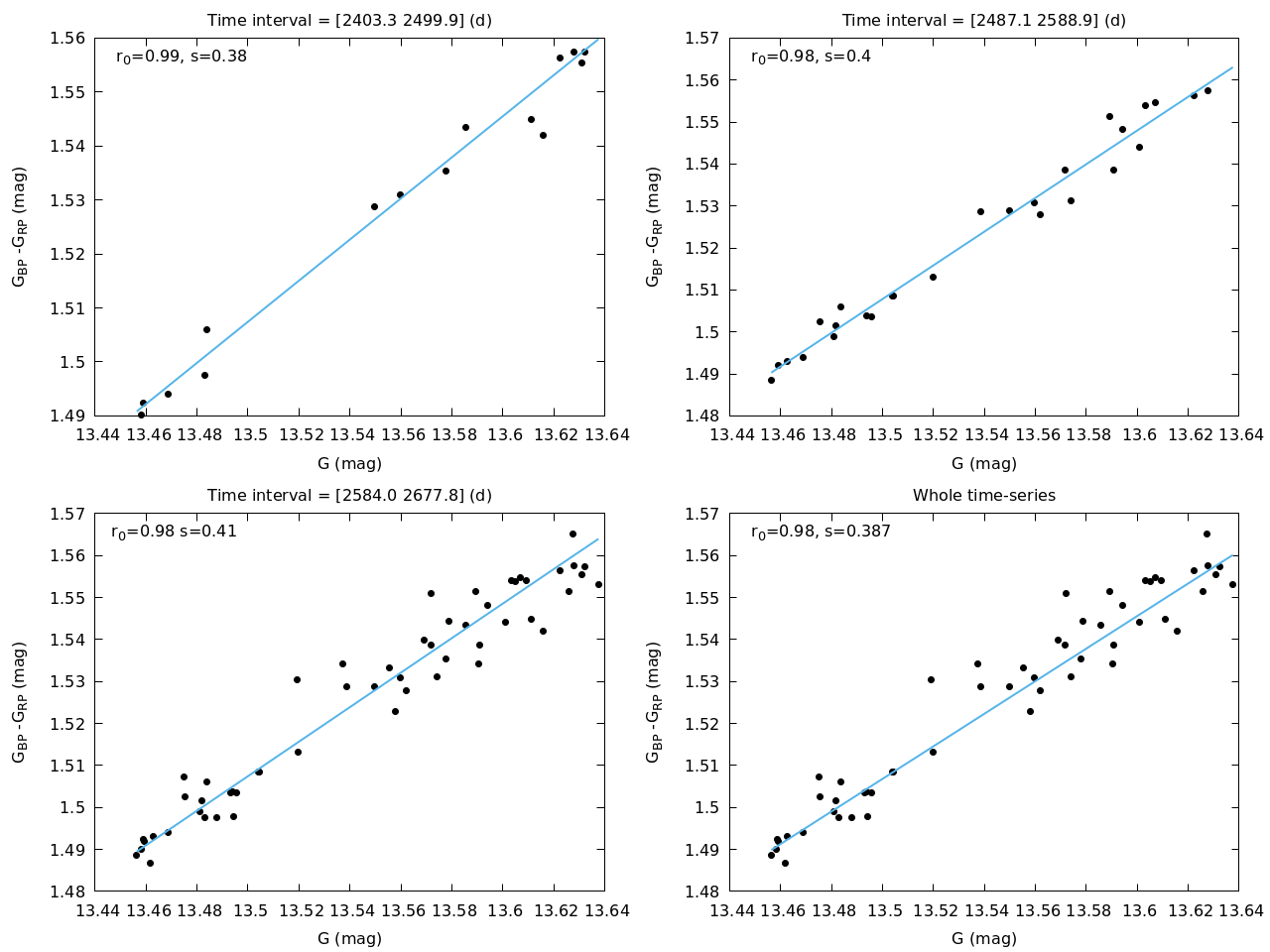}
\caption{$G~vs.~(G_{\rm BP}-G_{\rm RP})$ diagram for the star {\tt Gaia DR3 3321251347610387968}. In the first three panels we display the diagram for three distinct time-series segments whereas in the bottom-right panel we report the diagram for the whole time-series.}
\label{rcmc1}
\end{center}
\end{figure*}

In Figs. \ref{rcmc1}-\ref{umc} we report some examples of the different cases.

\begin{figure*}
\begin{center}
\includegraphics[width=160mm]{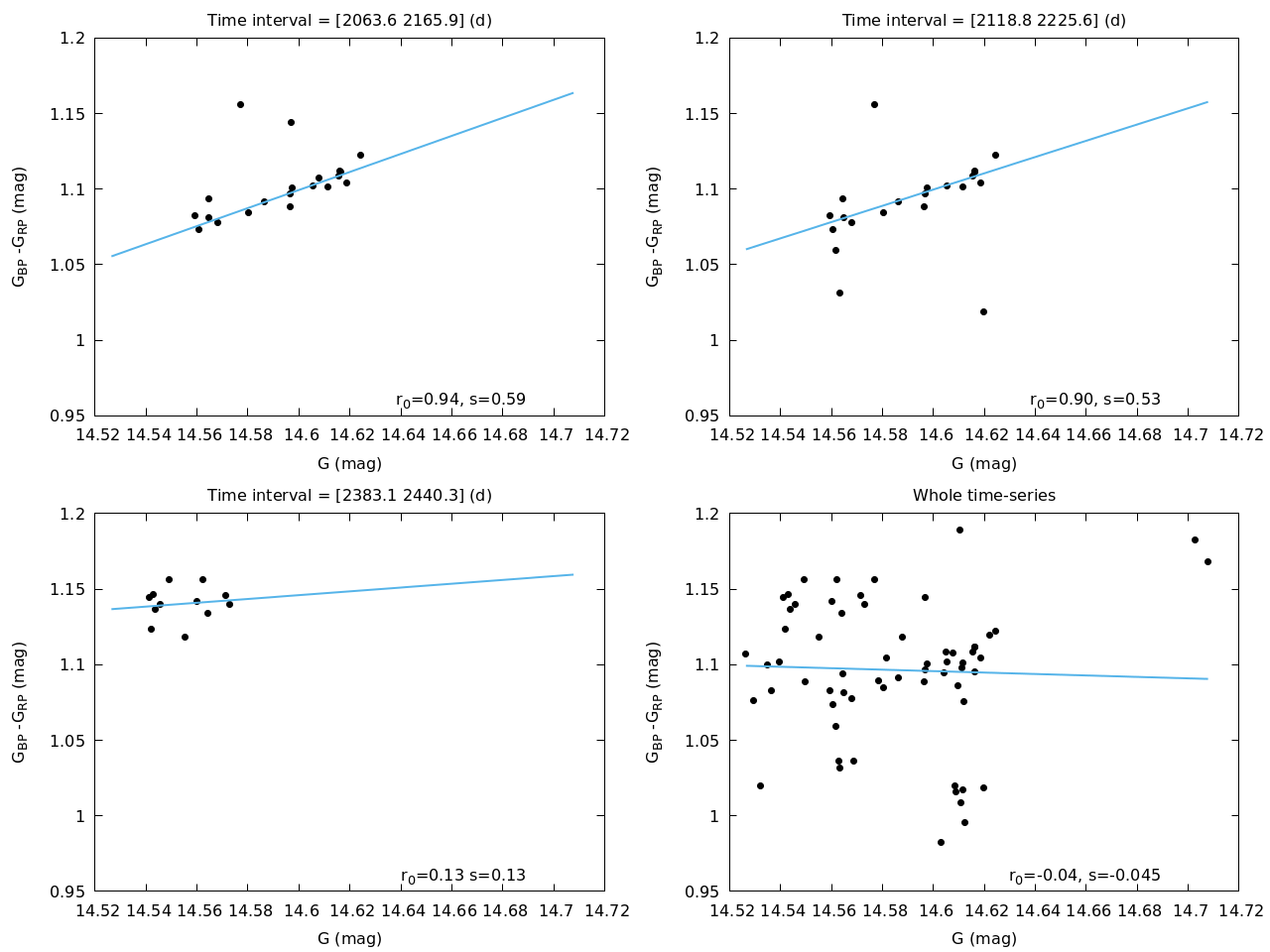}
\caption{Same of Fig. \ref{rcmc1} for the source {\tt Gaia DR3 6629372460509004928}. }
\label{rcmc2}
\end{center}
\end{figure*}
\begin{figure*}
\begin{center}
\includegraphics[width=160mm]{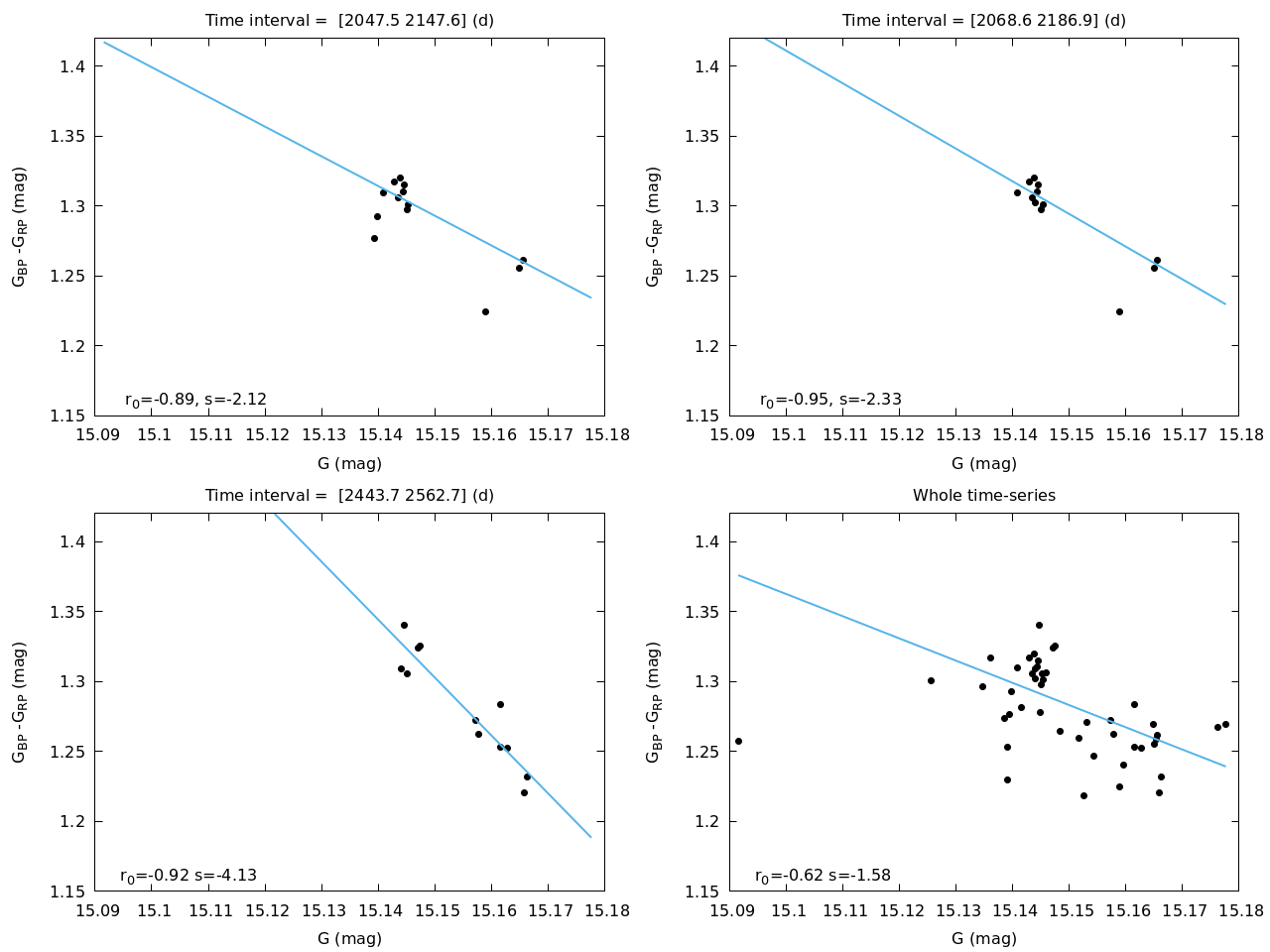}
\caption{Same of Fig. \ref{rcmc1} for the source {\tt Gaia DR3 5020821858561059456}. }
\label{bcmc}
\end{center}
\end{figure*}
\begin{figure*}
\begin{center}
\includegraphics[width=160mm]{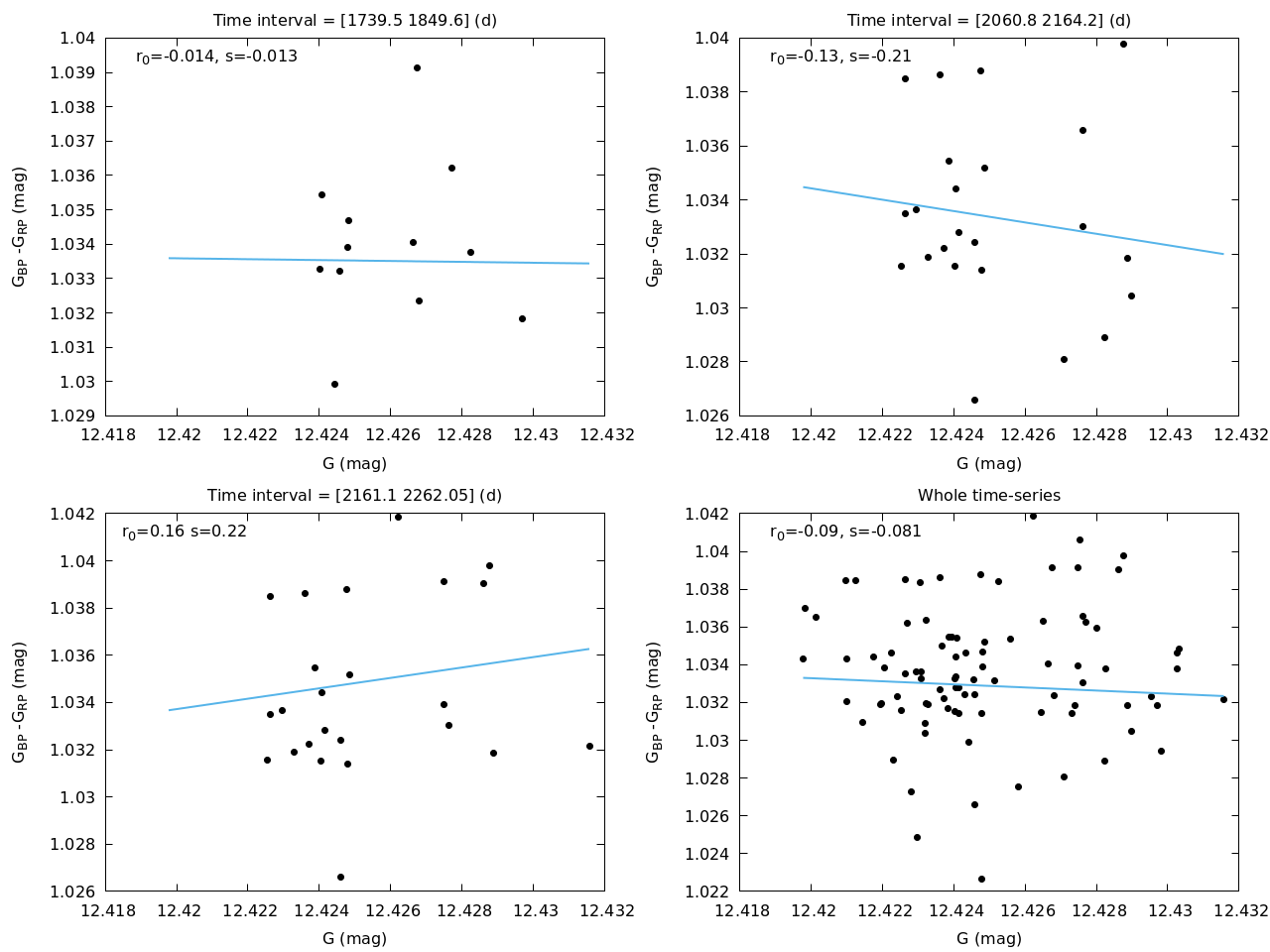}
\caption{ Same of Fig. \ref{rcmc1} for the source {\tt Gaia DR3 5823916073592266624}.}

\end{center}
\end{figure*}

In Fig. \ref{rcmc1} we illustrate an example of an RCMC star that is given by  the source {\tt Gaia DR3 3321251347610387968}. In the first three panels we display the $G~vs.~(G_{\rm BP} -G _{RP})$ diagram for three distinct segments whereas in the right-bottom panel we report the same diagram for the   whole time-series. In this example, the colour and brightness variations are strongly and significantly correlated ($r_0 > 0.9$ and $p < 0.05$) in each segment and  in the whole time-series.

In other RCMC stars, the positive correlation is seen only in some segments and is lost in the whole time-series because of the intrinsic evolution of spots and faculae. An example of this kind of star is given in Fig. \ref{rcmc2} where we plot the $G~vs.~(G_{\rm BP} -G_{\rm RP})$ diagram for the source {\tt Gaia DR3 6629372460509004928}. In this source, a significant correlation is observed only in the first two time-series segments. Such a correlation disappears in the third segment ($r_0 = 0.13$) and in the whole time-series ($r_0=-0.04$)

In Fig. \ref{bcmc} we illustrate an example of a BCMC star that is given by the source {\tt Gaia DR3 5020821858561059456}. In this star a significant anti-correlation is seen in all the three displayed segments ($r_0 \le -0.9$ and $p <0.05$). The anti-correlation is visible also in the whole time-series but a bit attenuated ($r_0 \le -0.62$ and $p < 0.05$).

Finally in Fig. \ref{umc} we illustrate an example of UMC star that is given by the source {\tt Gaia DR3 5823916073592266624}.

The values of the $r_0$, $p$, $s$ and $i$ parameters are complex functions of different factors like the stellar effective temperature $T_{\rm eff}$, the contrast in temperature $\Delta T_{\rm spot}$ between the spots and the stellar photosphere, the contrast in temperature $\Delta T _{\rm fac}$ between the faculae and the stellar photosphere, the geometrical distribution of spots and faculae and the photometric noise of the $G$, \bp and \rp time-series.
A full understanding of the $r_0$, $p$, $s$ and $i$ distribution seen in our sample of stars would require a theoretical modeling that is beyond the scope of this work. 
However, in the final section of the paper, we will show how $r_0$, $p$, $s$ and $i$ correlate with the other parameters listed in the {\tt gdr3\_rotmod} catalog, we will discuss how these relationships can shed light on the different nature of RCMC and BCMC stars and we will compare our results with previous studies.

\subsection{Search for the rotational modulation signal}
\label{sec:rotmodsignal}
The cleaned time-series segments are processed by means of the Lomb-Scargle algorithm as implemented by \cite{2009A&A...496..577Z}.  For each segment, the period $P$ corresponding to the highest peak of the periodgram is selected and its FAP (False Alarm Probability) is evaluated according to the \cite{2008MNRAS.385.1279B} prescriptions.
If the FAP value associated with $P$ satisfies the condition:
\begin{equation}
\label{fap}
FAP \le 0.05
\end{equation}
 then the period is flagged as valid and a sinusoidal model is fitted to the photometric  data:

\begin{equation}
\label{model}
m(t)=a_{m} + b_{m}\cos\left(\frac{2\pi t}{P}\right) +c_{m}\sin\left(\frac{2\pi t}{P}\right) 
\end{equation}

with $m \in$ ( \g, \bp, \rp). 
The FAP threshold given in Eq. \ref{fap}  has the aim to maximise the ratio between true and false positives and is based on the analysis of  \cite{2015MNRAS.450.2052S}. 
Note that while the fitting procedure is applied to all the three sets of photometric data, the period search is performed only on the \g  sub-series because  they have a higher Signal-To-Noise ratio (SNR).
After the fitting procedure, the residuals $\epsilon_i$ with respect to the fitted sinusoidal signals are computed. 
The transits with the higher residuals are flagged as  candidate flare events.
A given transit is flagged as outlier if it satisfies the condition:
\begin{equation}
|\epsilon_i|  \ge ~<|\epsilon|> +3\sigma_{|\epsilon|}  
\end{equation}

The amplitude of the rotational modulation signal is related to the non-axisymmetric part of the spot distribution and is often used as an index of the stellar magnetic activity level  \citep[see e.g.][]{2000A&A...358..624R,2015A&A...583A.134F,2013A&A...560A...4R,2016A&A...588A..38L}.
However the data coming from the {\it Kepler} mission showed that this index has to be treated with caution \citep[see e.g][]{2013A&A...560A...4R,2014ApJS..211...24M,2018ApJ...863..190B}. We will discuss the use of such an index in the next section, where the relation between the amplitude of the rotational modulation and the stellar rotation period is investigated.

\begin{figure}
\begin{center}
\includegraphics[width=80mm]{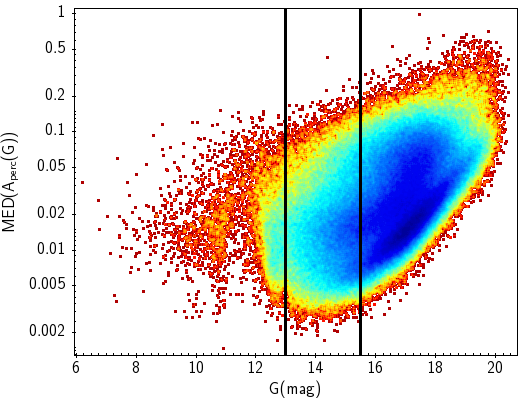}
\caption{Median photometric amplitude $A_{\rm perc}(G)$ vs. $G$ mag. }
\label{magai}
\end{center}
\end{figure}

The pipeline makes two different estimates of this index and provides a percentiles-based index\footnote{This index is very similar to the variability range defined by \cite{2010ApJ...713L.155B,2011AJ....141...20B} for the {\it Kepler} data. The only difference is that the {\it Kepler} index is computed in flux units instead of mag units.}:
\begin{equation}
\label{aper}
A_{\rm perc}(m)= m_{95\mathrm{th}} -m_{5\mathrm{th}},\end{equation}
and a fit-based index:
\begin{equation}
\label{afit}
A_{\rm fit}(m)= 2 \sqrt{b_m^2 +c_m^2}
\end{equation}
where $m_{95\mathrm{th}}$ and $m_{5\mathrm{th}}$ are the 95-th and 5-th percentile of the magnitude distribution in a given segment, respectively, and $b_m$ and $c_m$ are the coefficients of Eq. \ref{model} and $m \in$ ( \g, \bp, \rp).
So, overall, the pipeline computes six activity indexes for each segment in which a significant period $P$ ($FAP~\le~0.05$)  is detected. The amplitude inferred from the fit can be useful when dealing with faint sources. In fact, in faint sources the activity index computed with equation \ref{aper} can be over-estimated because of the increased photometric noise. The activity indexes in the \bp and \rp bands can be useful to constrain MARs temperature. In all the following analysis we will make use, for brevity, only of the $A_{\rm per}(G)$ index. Moreover we will mainly focus our analysis on stars with $G$ magnitude in the range (13,15.5), that is the interval where the \gaia photometric sensitivity reaches its maximum, as illustrated in Fig. \ref{magai}. In this picture, we reported, for each star, the median activity index $MED(A_{perc}(G))$ vs. the $G$ magnitude. The lowest amplitudes detected by the pipeline are of about 0.003 mag and fall in the $G$ range (13,15.5).

The best estimate for the stellar rotation period (supplied as {\tt best\_rotation\_period} in the {\tt gdr3\_rotmod} catalog) is computed by taking the mode  of the significant periods distribution
\citep[see][for details on the algorithm used to compute the mode.]{2018A&A...616A..16L,DR3-documentation}

\subsection{Spurious periods filtering}
\label{sec:prot}
As discussed in \cite{2018A&A...616A..16L}, the non-uniformity of the Gaia sampling could cause the detection of spurious periods. 
In these cases, the visual inspection of folded light curves reveals that  the data are poorly sampled and not well reproduced by the fitted sinusoidal curve. 
In order to reject these spurious periods, the FAP parameter needs to be complemented with other quality indicators.
We defined four quality assurance parameters that are the phase coverage (PC), the maximum phase gap  (MPG), the ratio $Q$ between the $A_{\rm fit}(\g)$ and the $A_{\rm perc}(\g)$ indexes and the reduced chi-squared $\tilde\chi^2(\g)$ associated with the fitting procedure described in Sec. \ref{sec:rotmodsignal}.

Each time-series segment, in which a significant period $P$ is detected, is folded according to $P$ and divided in 10 equally spaced phase bins.
The PC parameter is defined as the fraction of bins containing at least one data point.
The $MPG$ parameter is defined as
\begin{equation}
    MPG=max(\phi_i - \phi_{i+1})
\end{equation}
where $\phi_i$ is the phase of the i-th point of the folded light curve. 
The reduced chi-squared is defined according to the classical formulation:
\begin{equation}
\label{chi}
\tilde\chi^2(G)=\frac{1}{d}\sum_{i=1}^{N} \frac{( G_i- E_i)^2}{\sigma_i^2}
\end{equation}
where $N$ is the number of observations in the segment, $d=N-3$, $\g_i$  the i-th data-point of the segment, $\sigma_i$ the associated photometric error and $E_i$ the value predicted by the fitted sinusoidal model.

The $Q$ parameter is defined as:
\begin{equation}
\label{q}
Q=\frac{A_{\rm fit}(G)}{A_{\rm perc}(G)}
\end{equation}

The PC and MPG are used to establish the goodness of the sampling, whereas  the $\tilde\chi^2$ and $Q$ parameters are used to evaluate the goodness of the fit.
High PC values and small MPG values indicate a good sampling, whereas $\tilde\chi^2$ and $Q$ values close to 1 indicate a good consistency between the data and the fitted sinusoidal function.
The $Q$ parameter was introduced because a low chi-squared is not always sufficient to establish the consistency between the sinusoidal model and the data, as discussed below.

\begin{figure}
\begin{center}
\includegraphics[width=80mm]{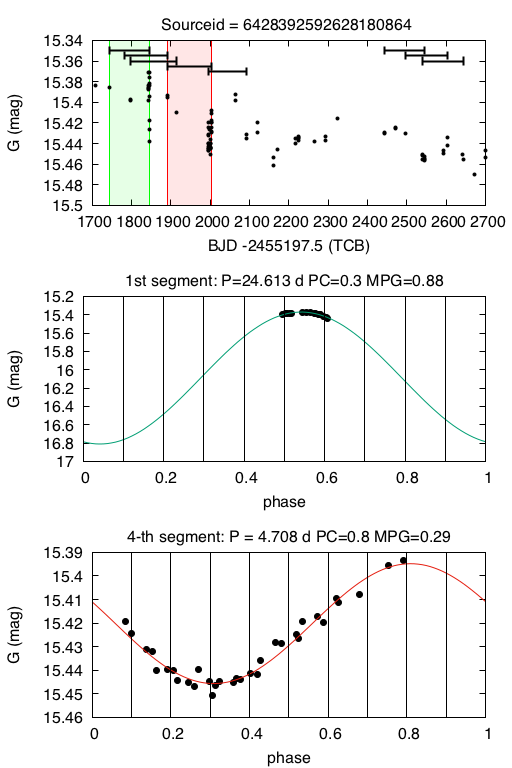}
\caption{Top panel: full $G$ time-series for the star {\tt Gaia DR3 6428392592628180864}. The dark segments enclose the sub-series extracted by the segmentation algorithm. The period search algorithm detected the periods $P=24.613~d$ and $P=4.708~d$ in the first and in the fourth segment, respectively. Middle panel: first sub-series folded according to the period $P=24.613~d$. Though the FAP associated with the period is below the rejection threshold, the period is discarded because of the low PC and high MPG values. Bottom panel: fourth sub-series folded according to $P=4.708~d$. In this case the detected period is flagged as valid because the sub-series exhibits an excellent phase coverage (PC=0.8) and a very small phase gap (MPG=0.29).   }
\label{phasecover}
\end{center}
\end{figure}

In Fig. \ref{phasecover}, we show the illustrative case  of the source {\tt Gaia DR3 6428392592628180864}. In the top panel, the whole \g time-series is plotted. The top black segments mark  the segments in which the time-series has been divided.
In the first segment  (high-lighted with the green shadow), the pipeline detected the period $P=24.613~d$ associated with a $FAP=3\times10^{-7}$ whereas in the fourth red shaded segment   the detected period is $P_4=4.708~d$ with a $FAP=3\times10^{-17}$. The   central and the bottom panel display the photometric points of the two segments folded according to the detected periods. The sinusoidal curves computed by the fitting procedure are over-plotted on the data. Both periods are associated with FAP values lower than the threshold used to flag a detection as reliable, but in the first segment the fitted curve is poorly sampled (PC=0.3)  and  the photometric points are mainly concentrated in only one phase bin (MPG=0.88).  
Moreover the visual inspection of the folded light curves reveals  that, in the first segment, though the reduced chi-square is quite close to 1 ($\tilde\chi^2=4.44$),  the fitted sine wave cannot be regarded as a reliable model for the data. The inconsistency between the data and the model, in this case, is reflected by the high $Q$ value. Indeed the variability  amplitude    inferred by the data distribution and that inferred from the fitted sine wave are noticeably different from each other with $A_{\rm perc}(\g)=0.054~mag$, $A_{\rm fit }(\g)=1.43~mag$ and $Q=26.4$. 

A given star is flagged and released as a Rotational Modulation variable if the segments used to infer the {\tt best\_rotation\_period} parameter, satisfy the following requirements:
\begin{enumerate}
\item{$0.5 \le Q \le 1.6$ in all  segments;}
\item{$PC \ge 0.4 \mbox{~and~} MPG <  0.3\mbox{~and~}\tilde\chi^2(\g) \le 32.5$ in at least one of the segments }
\item{the segment satisfying the requirement  (2) needs to be  different from the whole time-series.}
\end{enumerate}

The adopted threshold values occurring in the first requirement are the 5-th and 95-th percentile of the $Q$ distribution, respectively. The threshold value adopted in the second requirement is the 95-th percentile of the $\tilde\chi^2(\g)$ distribution.

The above mentioned criteria are more strict than those adopted in the DR2 release. Indeed in DR2 only the first and second requirement were adopted and the second requirement did not include the condition on $\tilde\chi^2(\g)$.
The third requirement was added because DR3 time-series span a longer time interval than DR2 time-series and, as showed in Sec. \ref{sec::segmentation}, the long-term evolution of MARs could seriously affect the detection of the correct rotation period.

\begin{figure}
\begin{center}
\includegraphics[width=80mm]{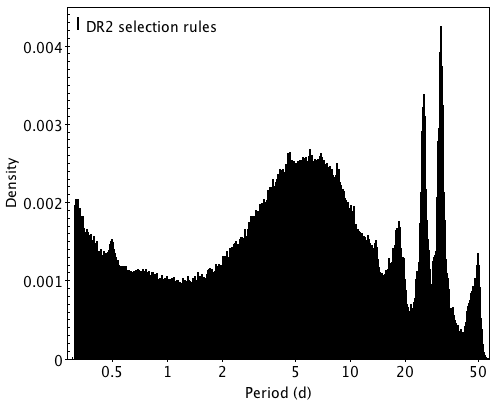}
\caption{Distribution of the rotation periods detected by the pipeline. The sharp peaks at 0.5, 18, 25, 32 and 49~d are spurious periods due to sampling and instrumental issues.	}
\label{olddensity}
\end{center}
\end{figure}

\begin{figure}
\begin{center}
\includegraphics[width=80mm]{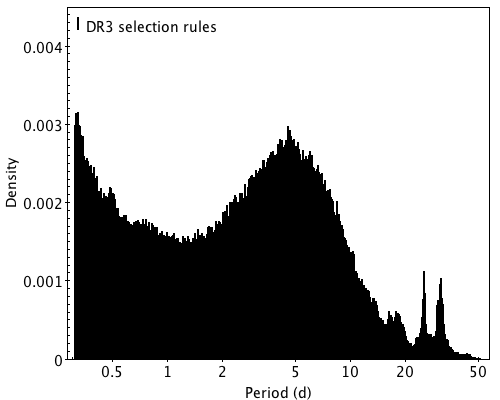}
\caption{Distribution of the rotation periods after the cleaning procedure described in Sec. \ref{sec:cleaning}. The peaks of spurious periods are remarkably flattened but are still persistent in the distribution.}
\label{newdensity}
\end{center}
\end{figure}

The necessity to enforce the DR2 requirements is  evident from the comparison between Figs.\ref{olddensity} and \ref{newdensity} where we plotted the {\tt best\_rotation\_period} distributions obtained by applying the DR2 and the DR3 requirements, respectively.  

\begin{figure}
\begin{center}
\includegraphics[width=80mm]{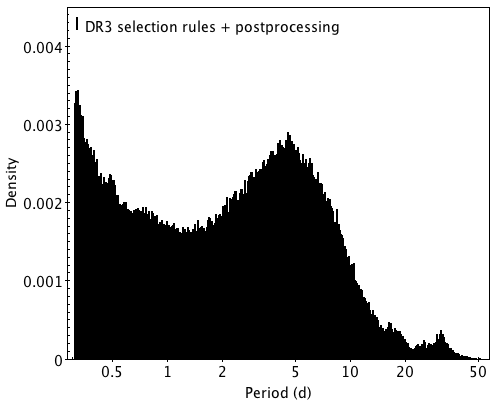}
\caption{Distribution of the rotation periods after the post-processing operations described in \ref{sec:ppf}.	}
\label{finaldensity}
\end{center}
\end{figure}

The use of the DR3 requirements filters out the most of the spurious periods centred around 0.5, 18, 25, 32 and 49~d.
Note that DR3 requirements are still not sufficient to remove all the spurious peaks of the distribution.
A further filtering was performed during the post-processing operations as discussed in Appendix \ref{sec:ppf}.  
These post-processing operations permitted to further clean the sample of the detected variables and to obtain a final catalogue of 474\,026 bona-fide stars with rotational modulation. The period distribution of this final sample of stars is reported in Fig. \ref{finaldensity}.

\section{Catalogue overview}
The {\tt gdr3\_rotmod} catalogue reports 474\'026 bona-fide stars with rotational modulation.
For each star it provides 66 different parameters whose full description can be found in the Sec. 3.3.15 of the GDR3 documentation \citep{DR3-documentation}.
In this section we show the correlation between some of these parameters and we discuss the impact of the catalogue on the understanding of the variability phenomena seen in magnetically active stars.

The completeness of the catalogue is limited due to the peculiarity of the \gaia scanning law and it is  about 0.4\% down to the limit magnitude $G\simeq21.5$  and about 4\% down to the limit magnitude $G\simeq15$. These latter are only average values and they can significantly change across the sky. In Appendix \ref{sec::completeness} and in Figs. \ref{completeness} and \ref{completeness1315} we show how the detection efficiency (that is an upper limit for the survey completeness) varies with the stellar magnitude and the ecliptic coordinates. The completeness is also a function of the stellar rotation period. In fact, \cite{2012MNRAS.421.2774D} demonstrated that the \gaia scanning law favors the detection of stars with short rotation periods (P < 5 d).

The contamination level of the catalogue has been assessed between 6 and 14 \% and is mainly due to binary systems (see details in  Sec. \ref{contamination}) The rate of correct period detection (estimated by the comparison between \gaia and other photometric surveys) has been assessed between the 70\% and the 80\%. The incorrect detections are partly due to the sparseness of the \gaia sampling and partly due to the physical properties of magnetically active stars (see details in App. \ref{correctdetections}) . 

Because of the low completeness, the catalogue  cannot be regarded as fully representative of the entire population of magnetically active stars. Nevertheless, as it will be discussed in this section, it is rich of precious information and details never seen by previous  surveys.

\subsection{Stable and unstable stars}
In Sec. \ref{sec::segmentation} we showed two different types of light curves (see Figs. \ref{example1}-\ref{example4}). In some  stars the rotational modulation signal is very stable whereas in others it looses coherence in time. This lost of coherence can be ascribed to different reasons: in some cases the instability could be due to the intrinsic evolution of MARs that induces a shift in the phase of the rotational signal, in other cases it could be induced by  a combined  effect of Surface Differential Rotation (SDR) and of MARs latitude migration \citep[][]{2016A&A...591A..43D} or,   (in most of the cases), it could be determined by the appearing and disappearing of MARs at random stellar longitudes  \citep{2020ApJ...901...14B}.

\begin{figure*}
\begin{center}
\includegraphics[width=160mm]{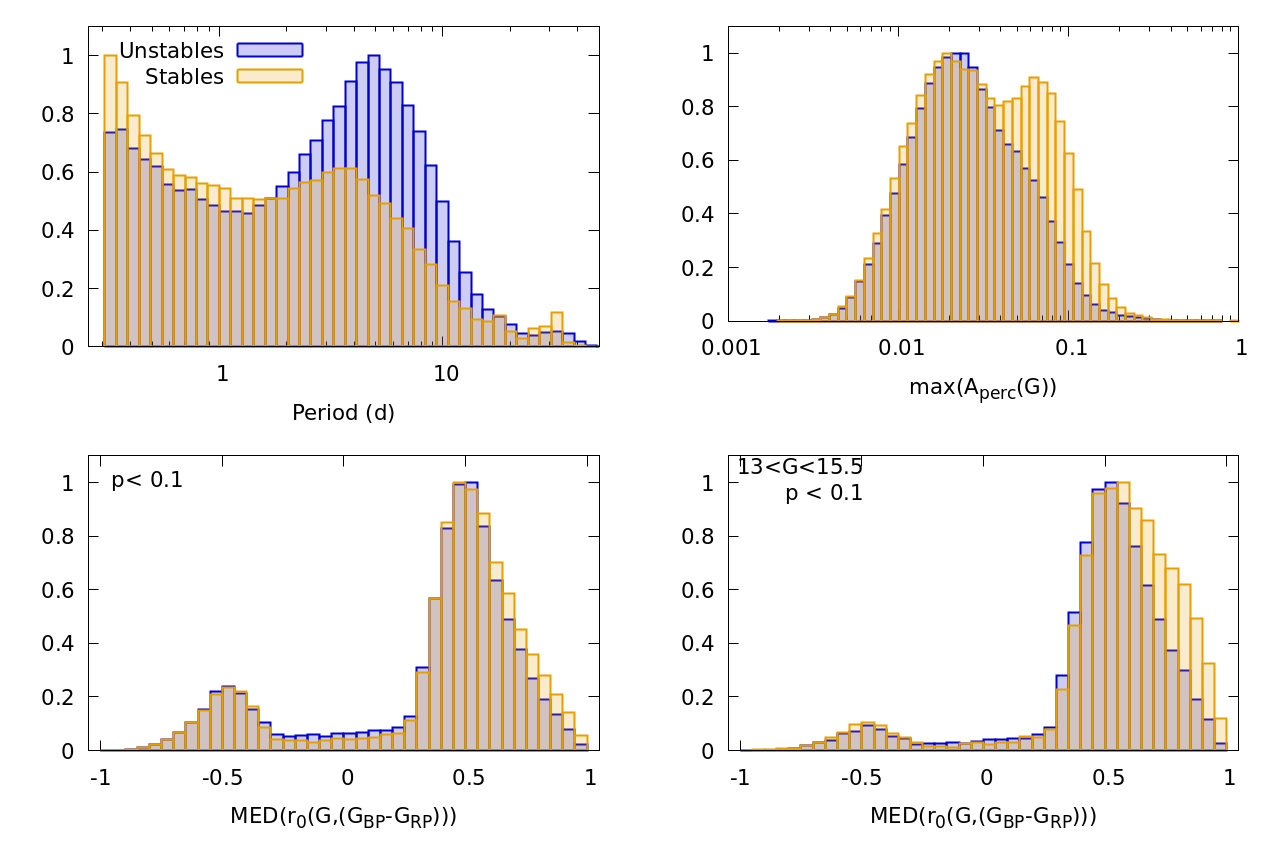}
\caption{Normalised distributions of stable and unstable stars.}
\label{stun}
\end{center}
\end{figure*}

We grouped the whole sample of rotational modulation variables in two distinct sets.
We flagged a  light-curve as stable if it satisfies two conditions:
\begin{itemize}
    \item {the period search algorithm was able to find a meaningful (FAP < 0.05) period in the whole time-series\footnote{the rotation period retrieved in the whole time-series is stored in the {\tt gdr3\_rotmod} catalogue and is given by the last element of the {\tt segments\_rotation\_period} array; the associated FAP is given by the last element of {\tt segments\_rotation\_period\_fap}}};
    \item {the period retrieved in the whole time-series differs less than 5 \% from the {\tt best\_rotation\_period} parameter}
\end{itemize}
The stars satisfying these conditions are the 38 \% of the whole sample. The remaining stars were flagged as unstable.
In Fig. \ref{stun} we show the normalised distributions of periods, amplitudes and $r_0$ values for both sets of stars. The stable stars have, on average, period shorter than the unstable  stars (see top left panel of the picture). This is in agreement with the findings of \cite{2022ApJ...924...31B}, who reported that $\tau_{{MARs}_{rot}}$ (i.e. the MARs life-time expressed in units of the stellar rotation period) tends to be longer for stars with shorter rotation periods. 
The amplitude distribution (top right panel) is bimodal for stable stars, with two peaks at about 0.015 and 0.005 mag, and unimodal for unstable stars, with a peak at about 0.015 mag. In the bottom left panel we reported the normalised distributions of $r_0$ values with $p<0.1$ for both sample of stars. The distribution of stable stars is slightly skewed towards positive values. This trend is more evident if we limit our analysis to stars falling in the magnitude range (13,15.5) where the photometric sensitivity of \gaia is higher (see Fig. \ref{magai}).

\subsection{Period-amplitude diagram}
In the present  section we treat the amplitude $A$ of the rotational modulation signal as a proxy of the stellar magnetic activity level and we investigate its relationship with the stellar rotation period $P$.

The relationship between the stellar magnetic activity 
and rotation period has been investigated in several papers.
According to the traditional picture, based on magnetic activity indicators like the coronal X-ray emission or the intensity of chromospheric emission lines, the level of magnetic activity increases towards shorter rotation periods and, depending on the stellar mass, saturates below a critical rotation period $P_{\rm crit}$ \citep[see e.g.][]{2003A&A...397..147P,2014ApJ...794..144R,2018MNRAS.476..908F}.

Different authors point out that the amplitude of the rotational modulation signal has a similar dependence on the stellar rotation period than other magnetic activity indicators  and that therefore this index can be used as a stellar activity proxy \citep[see e.g][]{2003A&A...410..671M,hartman,2018MNRAS.476.1224A}. However these works are based on limited sample of stars.   
In the last years the {\it Kepler} mission
permitted to investigate the relationship between this index and the stellar rotation period for several thousands of stars \citep[see e.g][]{2013A&A...560A...4R,2014ApJS..211...24M,2018ApJ...863..190B}.
The {\it Kepler} data confirmed that, despite a certain scatter, the amplitude $A$ tends actually to increase towards shorter rotation period and then to saturate below a limiting rotation period.

The  Gaia DR2 data  permitted to investigate with great detail the saturated regime of the $P-A$ diagram and revealed, for the first time,  the existence of a family of fast rotating ($P ~ \le ~ 2~d$) stars with small amplitudes.

\begin{figure}
\begin{center}
\includegraphics[width=80mm]{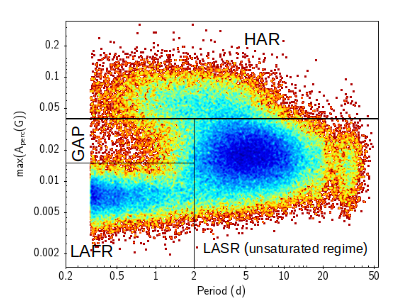}
\caption{$P-A$ density diagram for the {\tt gdr3\_rotmod} stars with \g $mag$ falling in the (13,15.5) interval.}
\label{padr3}
\end{center}
\end{figure}

Such a feature has been confirmed by the DR3 data.
In Fig. \ref{padr3} we display the $P-A$ diagram for  the stars of the {\tt gdr3\_rotmod} catalogue with a \g magnitude falling in the interval (13,15.5).  We restricted our analysis to such an interval because at brighter and fainter magnitudes the photometric uncertainty could prevent the detection of the low amplitude signals and, consequently, bias the amplitude distribution (see Fig. \ref{err}).

As outlined in Sec. \ref{sec:rotmodsignal},
for a given star the catalogue reports different values of the $A_{\rm perc}(G)$ index (one for each of the segments extracted from the whole time-series).
The amplitude used to obtain Fig. \ref{padr3} is given by the maximum value of these activity indexes i.e. by $max(A_{\rm perc(G)})$ \footnote{this value is stored in the tt{gdr3\_rotmod} catalogue for each star and is given by the index {\tt maximum\_activity\_index}}. The rotation period adopted to build the plot is given by the {\tt best\_rotation\_period} parameter described in Sec. \ref{sec:rotmodsignal}. The points of the $P-A$ diagram are colored according to the target density. The blue regions are those with a higher density whereas the yellow ones are those with a lower density.

The picture clearly shows three different clusters of rotating stars that, according to \cite{2019ApJ...877..157L}, we call 
 High-Amplitude-Rotators  (HAR), Low-Amplitude-Fast-Rotators (LAFR) and Low-Amplitue-Slow-Rotators (LASR). 
More specifically the HAR branch is  the region satisfying approximately the condition:
\begin{equation}
    \label{HAFR}
    max(A_{\rm perc}) > 0.04 mag,
\end{equation}
 the LAFR branch is the area satisfying the condition:
\begin{equation}
    \label{lafr}
    P \le 2 d  \mbox{~and~} max(A_{\rm perc}) < 0.015 mag
\end{equation}

and the LASR branch is the area satisfying the  condition
\begin{equation}
    \label{lasr}
    P > 2 d \mbox{~and~} max(A_{\rm perc}) < 0.04 mag.
\end{equation}
The LAFR  and the HAR regions are, in particular,  well separated by a sparsely populated area that, in \cite{2019ApJ...877..157L}, is referred to as the gap region  and that has been interpreted  as the evidence of a rapid transition between magnetic configurations characteristic of HARs to that characteristic of LAFRs.

As mentioned above, the regime of the fast rotation is poorly explored by {\it Kepler}. The samples of stars analysed, for instance by \cite{2013A&A...560A...4R} and \cite{2014ApJS..211...24M} contain only a few hundred stars with periods shorter than 1 d, whereas the \gaia variables with period shorter than 1 d are about 150\,000 \footnote{We remind that, though the average detection efficiency of our pipeline is very low, it dramatically increases towards shorter rotation periods as demonstrated in \cite{2012MNRAS.421.2774D})}.
The comparison between the {\tt Kepler} and \gaia results performed by \cite{2019ApJ...877..157L} shows that, actually, the two sets of data are complementary and that the LASR family seen by \gaia is the "tip" of the unsaturated regime sampled by {\tt Kepler} \citep[see Fig. 5 of][for details.]{2019ApJ...877..157L}  
The LAFR branch, in {\tt Kepler} data, is not visible and it  has been highlighted, for the first time, by the \gaia.

\begin{figure}
\begin{center}
\includegraphics[width=80mm]{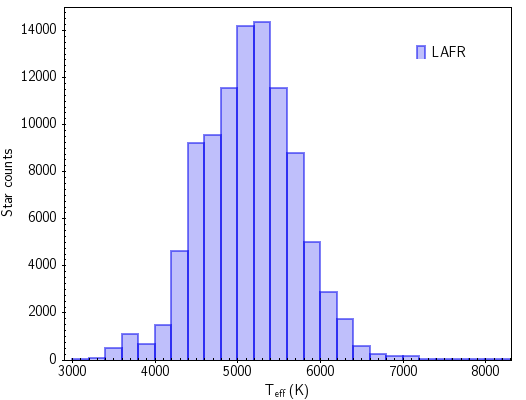}
\caption{$T_{eff}$ distribution for the LAFR stars.}
\label{lafrhisto}
\end{center}
\end{figure}

Recently, \cite{2021ApJS..255...17S} found also, in \textit{Kepler} targets, a few of fast rotating stars  with a low amplitude signal but they pointed out that these stars are all above the Kraft break \citep{1967ApJ...150..551K} i.e. stars characterized by an effective temperature a $T_{\rm eff} > 6200 K$.
This is not the case of the LAFR stars listed in the {\tt gdr3\_rotmod} catalog. Indeed, in Fig. \ref{lafrhisto} we show the $T_{\rm eff}$ distribution for all the LAFR stars. Almost all the sample of LAFR lie under the Kraft break. Note that the $T_{\rm eff}$ values used to build the distribution are the DR3 values inferred by the \bp and \rp mean spectra according to the procedure described in \cite{gaiaparams}.

The different intensity and topology of surface magnetic fields in  the three classes of rotating stars is still unknown, however a deep analysis of Gaia DR3 data can provide precious insight into their properties.

\begin{figure}
\begin{center}
\includegraphics[width=80mm]{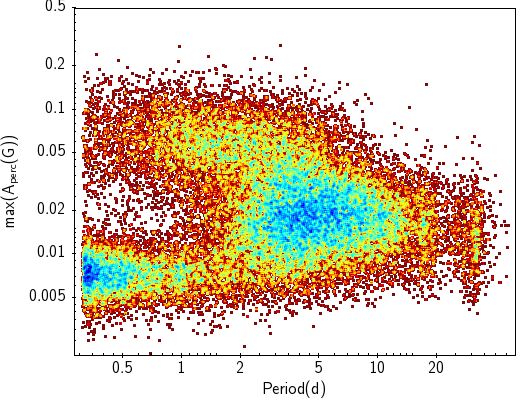}
\caption{$P-A$ density diagram obtained for the stable stars with $G$ mag values included in the \g range (13,15.5).}
\label{padr3stable}
\end{center}
\end{figure}

\begin{figure}
\begin{center}
\includegraphics[width=80mm]{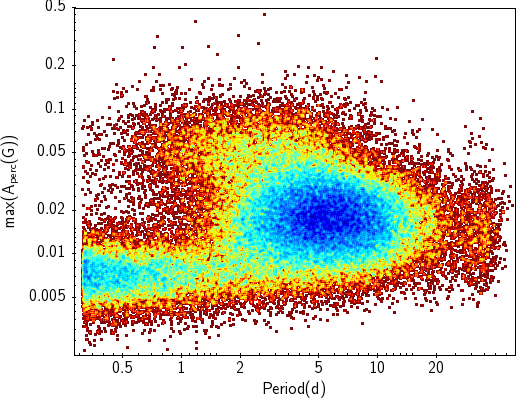}
\caption{$P-A$ density diagram obtained for the unstable stars with $G$ mag values included in the \g range (13,15.5). }
\label{padr3instable}
\end{center}
\end{figure}

In Fig. \ref{padr3stable} and \ref{padr3instable} we present the density $P-A$ diagram for the stable and unstable stars, respectively. In both pictures we used stars with $G~mag$ falling in the $(13,15.5)$ range for the reasons discussed at the beginning of this section.
The comparison between the two pictures shows that the stable stars are mainly concentrated in the HAR branch whereas the  unstable stars are mainly concentrated in the LAFR and in the LASR branches.
In Table \ref{stablestat} we reported for each branch the percentage of stable and unstable stars. The highest percentage of stable stars (47\%) is found in the HAR branch. This percentage  decreases to 34\% in the LAFR branch  and drops to 28\% in the LASR branch.   

\begin{table}
\label{stablestat}
\caption{Percentage of stable and unstable stars in the different regions of the $P-A$ diagram.}
\begin{tabular}{|l|l|l|}
\hline
  \multicolumn{1}{|c|}{Branch} &
  \multicolumn{1}{c|}{Stable} &
  \multicolumn{1}{c|}{Unstable} \\

\hline
   HAR & 47\% & 53\% \\
   GAP & 39\% & 61\% \\
   LAFR & 34\% &66\% \\
   LASR & 28\% & 72\%\\

\hline\end{tabular}
\end{table}

\subsection{The $P-A-r_{0}$ diagram.}
\begin{figure}
\begin{center}
\includegraphics[width=80mm]{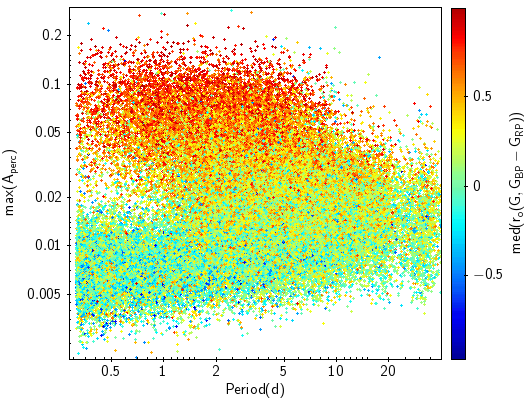}
\caption{$P-A$ diagram obtained for the stars with $G$ mag values included in the \g range (13.5,15). The colors of the points are coded according to $med(r_0((G_{\rm BP}-G_{\rm RP}), G))$ i.e. the median value of the Pearson correlation coefficients measured in the different segments.}
\label{padr3corr}
\end{center}
\end{figure}

Another important insight into the difference between the three branches is given by the analysis of $r_0(G, G_{\rm BP} -G_{\rm RP})$ distribution.
In Fig. \ref{padr3corr} we report the $P-A$ diagrams with data points color-coded according to the median value of the Pearson correlation coefficients obtained in the different segments i.e. $MED(r_0(G, G_{\rm BP} -G_{\rm RP}))$.
The picture has been obtained by selecting stars with magnitude $G$ falling in the interval (13,15.5). 
The visual inspection of the picture shows that all HAR stars  are also RCMC stars i.e. these stars exhibit a  strong positive correlation between magnitude and colourvariations. This correlation persists but is  attenuated in the LASR branch. Finally in the LAFR branch the colour and brightness variations are poorly correlated.

\begin{figure}
\begin{center}
\includegraphics[width=80mm]{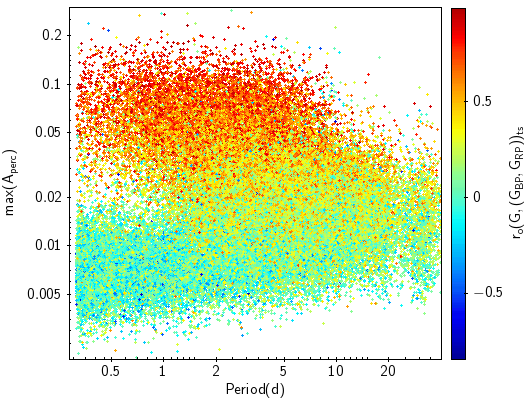}
\caption{$P-A$ diagram obtained for the stars with $G$ mag values included in the \g range (13.5,15). The colors of the points are coded according to $(r_0((G_{\rm BP}-G_{\rm RP}), G)_{TS})$ i.e. the value of the Pearson correlation coefficient measured in the whole time-series.}
\label{padr3corrts}
\end{center}
\end{figure}

The different nature of the three regimes of rotating stars is also evident in Fig. \ref{padr3corrts} where the points of the $P-A$ diagram are colored according to the Pearson Coefficient $r_{0_{TS}}$ value measured in the whole time-series. In HAR stars the correlation between magnitude and colour variation tends to be strong not only in the single segments but also in the whole time-series, indicating that in these stars the configuration of MARs is stable for the time interval covered by the Gaia DR3 time-series. The correlation tends to fade in the LASR branch and is close to 0 in the LAFR region.

The $A-P-r_0$ diagrams displayed in Figs. \ref{padr3corr} and \ref{padr3corrts} have been obtained without taking into account the statistical significance associated with $r_0$ otherwise the uncorrelated stars would have been erased from the plot. However these diagrams do not permit to  see the position of the small percentage of blueing stars present in the {\tt gdr3\_rotmod} catalogue(see Fig. \ref{histor0sig}).

\begin{figure}
\begin{center}
\includegraphics[width=80mm]{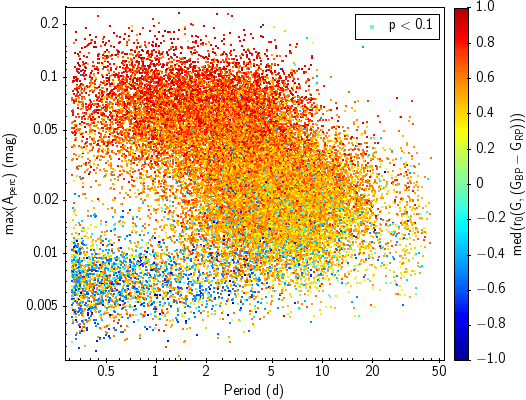}
\caption{Same of Fig. \ref{padr3corr} but for stars with $p < 0.1$.}
\label{padr3corr01}
\end{center}
\end{figure}

\begin{figure}
\begin{center}
\includegraphics[width=80mm]{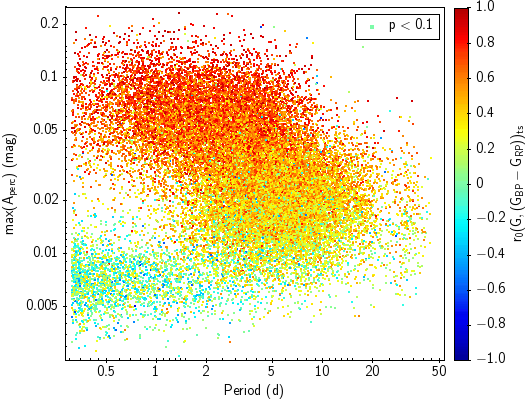}
\caption{Same of Fig. \ref{padr3corrts} but for stars with $p < 0.1$.}
\label{padr3corrts01}
\end{center}
\end{figure}

In Figs. \ref{padr3corr01} and \ref{padr3corrts01}, we displayed the same $A-P-{r_0}$ diagrams for the sample of stars in which the $p$ value associated with the median Pearson coefficient is less than 0.1.
In Fig. \ref{padr3corr01} the points are color-coded according to median Pearson coefficient whereas in Fig. \ref{padr3corrts01} according to the Pearson coefficient measured for the whole time-series. The visual inspection of the two pictures shows that blueing stars are mainly located in the LAFR branch and that the anti-correlation is observed only in the segments but not in the whole time-series where the Pearson coefficient is close to 0. The pictures clearly show that while the reddening stars located in the HAR region of the diagram are very stable in time (at least in the time spanned by the DR3 time-series), the blueing stars are quite unstable. 

As stressed in Sec. \ref{sec:colorbrightness}, the values of $r_o(G,(G_{\rm BP} -G_{\rm RP}))$ and $r_0(G_{\rm BP}, G_{\rm RP})$ are complex functions of the geometrical distribution of MARs and of the contrast between their temperature and the stellar photosphere temperature. 

A full understanding of the $A-P-r_0$ diagrams would require a theoretical modelling that will be the topic of a future paper (Distefano et al. 2022, in preparation). However some hint on their interpretation could be given by the works of \cite{2008A&A...480..495M} and \cite{2019ApJ...879..114I} that analysed patterns of brightness-colour variations in a sample of 14   magnetically active close binary stars and in a sample of about 12000 stars in the OGLE \citep[Optical Gravitational Lensing Experiment][]{2015AcA....65....1U} Galactic Bulge fields, respectively.

The two works have a slightly different approach because \cite{2008A&A...480..495M} splits the long-term time-series in intervals where the rotation modulation signal is stable whereas \cite{2019ApJ...879..114I} measures the correlation coefficient in the whole  OGLE time-series that span up to 13 years of observations. Hence the $r_0$ value computed by \cite{2008A&A...480..495M} is comparable to our $MED(r_0)$ index and the $r_0$ computed by \cite{2019ApJ...879..114I}  is more similar to our $r_{0_{ts}}$ value.

\cite{2008A&A...480..495M} interprets reddening variables as stars dominated by cool spots and uncorrelated variables as stars dominated by faculae. In particular he ascribes the lack of correlation between magnitude and colour to the fact that faculae are spatially or temporally uncorrelated with spots.
This for instance happens in our Sun where the typical life-time of faculae is longer than spots life-time \citep[see e.g.][]{2004A&A...425..707L}.
Such an interpretation and the location of reddening and uncorrelated stars in the $P-A-r_0$ diagram are coherent  with the works of \cite{2017ApJ...851..116M} and \cite{2018ApJ...865..142B} according to which spot-dominated stars have a variability amplitude higher than faculae-dominated stars. In particular, \cite{2018ApJ...865..142B} points out that faculae domination can occur also in fast rotating stars and that this is really surprising because theoretical models predict that fast rotators should be spot-dominated \citep{2014A&A...569A..38S}. 

As mentioned above, \cite{2019ApJ...879..114I} focus their analysis on the long-term correlation between brightness and colour variations. They interpret reddening variables as stars dominated by long-lived spots and uncorrelated variables as stars characterized by a high level magnetic activity and by a rapid change of MARs configuration. This is coherent with the bottom-right panel of Fig .\ref{stun}, according to which stable stars tend to have a higher correlation coefficient than unstable stars. 

\begin{figure*}
\begin{center}
\includegraphics[width=160mm]{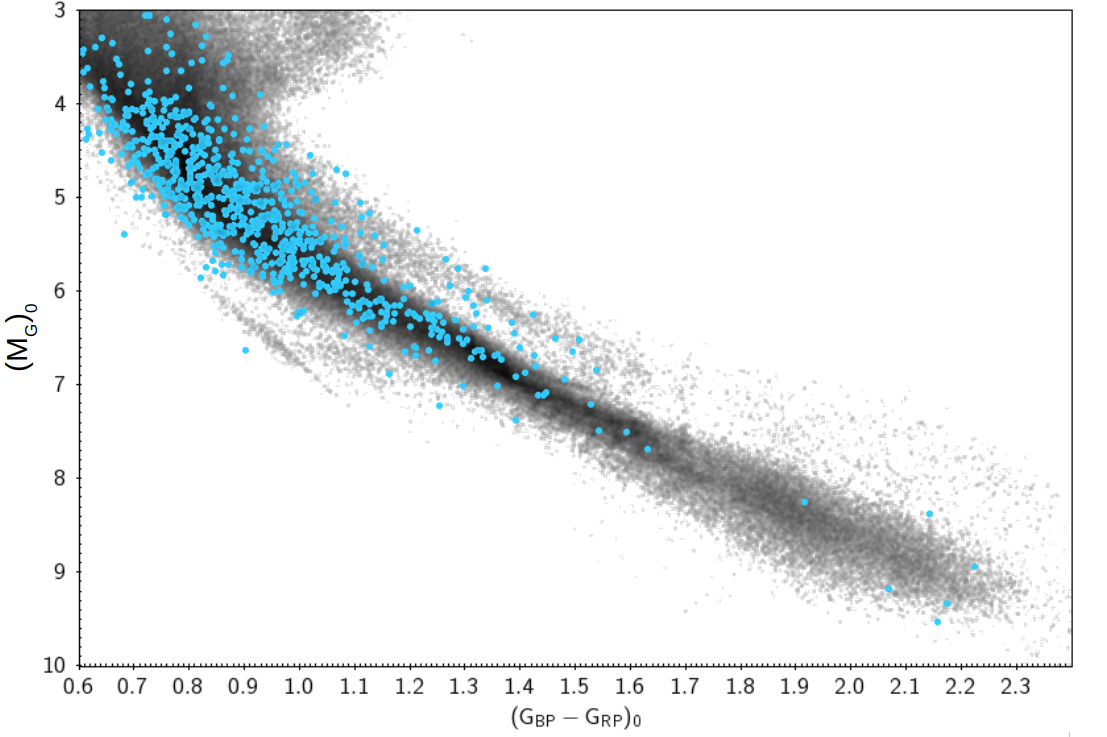}
\caption{$m_G~vs.(G_{\rm BP} - G_{\rm RP})$ diagram observed by \gaia. The blue bullets mark the location of the blueing variables. The region parallel to the main-sequence is the binary-sequence.}
\label{lumteff}
\end{center}
\end{figure*}

\begin{figure*}
\begin{center}
\includegraphics[width=160mm]{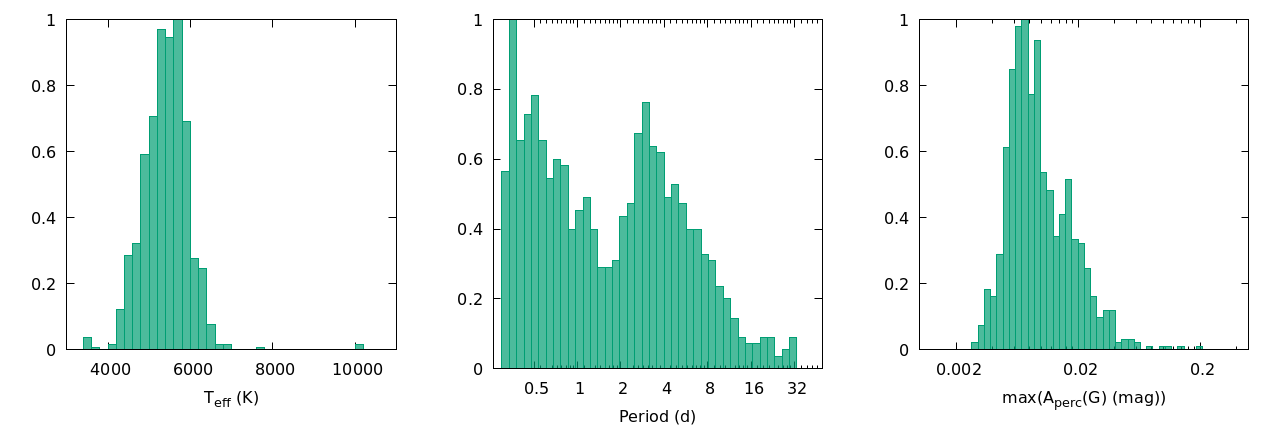}
\caption{Left panel:$T_{\rm eff}$ distribution for blueing stars. Central panel: rotation period distribution for blueing stars. Right panel: amplitude variability distribution for blueing stars. }
\label{bluehisto}
\end{center}
\end{figure*}

Finally, the meaning of blueing variables is quite puzzling.  
\cite{2008A&A...480..495M} interprets these stars as  binary systems characterized by a magnetically active  component and by an inactive earlier type stellar companion, whereas \cite{2019ApJ...879..114I} state that these variables could be peculiar stars covered by chemical spots characterized by an overabundance of heavy elements  and  by variable line-blanketing effect. However both explanations are unsatisfactory. Indeed, in Fig. \ref{lumteff} we reported the location of the blueing variable in the $(M_G)_0vs.(G_{\rm BP} -G_{\rm RP})_0$ diagram observed by \gaia, where $(M_G)_0$ and  and $(G_{\rm BP} -G_{\rm RP})_0$ are the absolute magnitude in the $G$ band and the colour $(G_{BP}-G_{RP})$ corrected for the interstellar reddening, respectively.
If blueing stars were binary systems, they should lie in the binary sequence of the diagram and this happens just for a few of them. 
Finally, if blueing variables were peculiar stars their spectral type should be earlier than $F8$ \citep{2019MNRAS.483.2300S} and their effective temperatures $T_{\rm eff}$ higher than $\simeq 6300 K$ and this happens just for a very small fraction of them (see left panel of Fig. \ref{bluehisto}).

\cite{rotorcct} found also an anti-correlation between colour and magnitude variations in a group of 5 Classical T-Tauri stars. They speculate that this pattern could be ascribed to the combined effect of the stellar occultation by the circumstellar disk and the inhomogeneous structure of the disk itself (see \cite{bouvier99} for a phenomenological  description  and \cite{bertout} for a theoretical model of this  variability phenomenon, respectively). However the variability amplitude associated with  this kind of events is usually of the order of tenths of magnitude   \citep[see e.g. light curves in][]{bertout,stauffer} whereas the blueing variables reported here have smaller amplitudes   (see second  panel of Fig. \ref{bluehisto}). Moreover the period distribution of the blueing variables exhibits a considerable fraction of short rotation periods (see second panel of Fig. \ref{bluehisto}) whereas the period distribution of  stars surrounded by a circum-stellar disk is usually skewed towards longer values (see, for instance, Fig.11 of \cite{2014MNRAS.444.1157D} and Fig. 8 of \cite{2018AJ....155..196R}).

\section{Conclusion}
In the present paper we described a method useful to analyse the Gaia photometric time-series and to spot possible spurious signals induced by the instrument or by the physical properties of the investigated source. This method, based on the statistical parameter $r_\text{exf}$ described in App. \ref{sec:ppf}, can be applied to all the \gaia DR3 photometric  time-series and can be useful to researchers interested in specific sources or photometric data. The values of  $r_\text{exf}$ will be published in the \gaia DR3 auxiliary table {\tt vari\_spurious\_signals} with the field name {\tt g\_spearman\_corr\_exf}  and will be available for all the released \gaia DR3 time-series.

The {\tt gdr3\_rotmod} catalog
reports 474\,026 magnetically active stars among which about 430\,000 stars are, as far as we know, new discovered variables. As discussed  in App. \ref{qualityassesment}, the completeness of this all-sky catalogue is about 0.4\% down to the limiting magnitude $G=21.5$ and about 4\% down to the limiting magnitude $G=15$. Such a low completeness is partly due to the \gaia survey properties (i.e. the time-series sampling and the technical issues discussed in App. \ref{qualityassesment} ) and partly due to the intrinsic nature of solar-like variables. In fact stars active or less active than the Sun, in which $\tau_{MARs}$ is shorter than a stellar rotation period, are hardly detectable by period search algorithms. For all these reasons the {\tt gdr3\_rotmod} catalogue cannot be considered as fully representative of the entire solar-like stars population. Nevertheless, this catalogue allows us to retrieve new and precious information on this class of variable stars, especially in the fast rotation regime, that is poorly explored by previous surveys. Indeed, about 150\,000 stars of the catalogue have rotation periods shorter than 1 day.

For each star of the catalogue we provided a list of about 70 parameters whose detailed description can be found in \cite{DR3-documentation}.

In the present paper we focussed on the stability of the stellar light curves and on the relationship between the period $P$, the amplitude $A$ of the rotational modulation signal and the Pearson Correlation Coefficient $r_0$ between the brightness and colour variations of the stellar sources.

The analysis of the $P-A$ diagram confirms the DR2 findings i.e. the existence of three different branches.
The LASR and the HAR branch correspond to the tip of the unsaturated regime and to the saturated regime, respectively. The LAFR branch is a new family of stars never revealed by previous surveys.

The stars located in the HAR branch tend to have a rotational modulation signal more stable in time than that
revealed in the LAFR  and in the LASR branches.

The analysis of the $P-A-r_0$ diagrams shows that in HAR  stars  there is a strong correlation between colour and brightness variations  and that such a  correlation is stable along the full 34-months Gaia time-series.

In LAFR stars, instead, the colour and magnitude variations tend to be uncorrelated.

According to the meaning attributed to $r_0$ by previous works, we can state that HAR stars are dominated by long-lived dark spots whereas LAFR stars are dominated by bright faculae and are characterized by rapidly variable magnetic fields.

The LASR stars are characterized by a moderate correlation between brightness and colour variations. This implies that dark spots are still the main cause of their variability but that faculae spatially or temporally uncorrelated with the spots tend to attenuate the correlation between magnitude and colour variations.

Finally there is a small fraction of stars in which the magnitude and colour variation are anti-correlated.
These stars are mainly concentrated in the LAFR and in the LASR  regions of the $P-A$ diagram.
As discussed Sect. 3.3, the interpretation of these variables is quite puzzling.
Accordingly to the literature these stars could be binary systems characterized by an active component and an inactive blue companion or  stars covered by chemical spots, but we showed  that their luminosity and effective temperature are incompatible with both explanations.  
A third explanation could be that the variability of these stars is driven by the variable extinction induced by an inhomogeneous circumstellar disk. However, such an hypothesis, discussed in the above section, also presents some criticism and should be tested with the aid of circum-stellar disk indicators. 

Despite the limitations due to the sparse sampling and to the instrumental issues discussed in Sects. \ref{sec:ppf} and \ref{dr2dr3}, \gaia DR3 provides the largest ever catalogue of rotational variables and permits to study the patterns of color-brightness variations from a statistical point of view.
The information coming from three photometric bands sheds light on the physical nature of the different branches seen in the period-amplitude diagram.

Such a  picture will be enriched in the future \gaia releases where all the issues discussed in Sects. \ref{sec:ppf} and \ref{dr2dr3} will be fixed and where the flaring analysis will be added.

\section*{Acknowledgements\label{sec:acknowl}}
\addcontentsline{toc}{chapter}{Acknowledgements}
The authors are very grateful to the referee Gibor Basri for the constructive comments and suggestion which remarkably helped to improve the  quality of the paper.

This work presents results from the European Space Agency (ESA) space mission \gaia. \gaia\ data are being processed by the Institutions participating in the \gaia\ MultiLateral Agreement (MLA). The \gaia\ mission website is \url{https://www.cosmos.esa.int/gaia}. The \gaia\ archive website is \url{https://archives.esac.esa.int/gaia}.
The \gaia\ mission and data processing have financially been supported by:
\begin{itemize}
    \item the Agenzia Spaziale Italiana (ASI) through contracts I/037/08/0, I/058/10/0, 2014-025-R.0, 2014-025-R.1.2015, and 2018-24-HH.0 to the Italian Istituto Nazionale di Astrofisica (INAF), contract 2014-049-R.0/1/2 to INAF for the Space Science Data Centre (SSDC, formerly known as the ASI Science Data Center, ASDC), contracts I/008/10/0, 2013/030/I.0, 2013-030-I.0.1-2015, and 2016-17-I.0 to the Aerospace Logistics Technology Engineering Company (ALTEC S.p.A.), INAF, and the Italian Ministry of Education, University, and Research (Ministero dell'Istruzione, dell'Universit\`{a} e della Ricerca) through the Premiale project `MIning The Cosmos Big Data and Innovative Italian Technology for Frontier Astrophysics and Cosmology' (MITiC);
    \item the Swiss State Secretariat for Education, Research and Innovation through the ``Activit\'{e}s Nationales Compl\'{e}mentaires’'.
\end{itemize}
The \gaia\ project and data processing have made use of:
\begin{itemize}
\item the Set of Identifications, Measurements, and Bibliography for Astronomical Data \citep[SIMBAD,][]{2000A&AS..143....9W}, the `Aladin sky atlas' \citep{2000A&AS..143...33B,2014ASPC..485..277B}, and the VizieR catalogue access tool \citep{2000A&AS..143...23O}, all operated at the Centre de Donn\'{e}es astronomiques de Strasbourg (\href{http://cds.u-strasbg.fr/}{CDS});
\item the software products \href{http://www.starlink.ac.uk/topcat/}{TOPCAT}, \href{http://www.starlink.ac.uk/stil}{STIL}, and \href{http://www.starlink.ac.uk/stilts}{STILTS} \citep{2005ASPC..347...29T,2006ASPC..351..666T};
\item Matplotlib \citep{Hunter:2007};
\item Astropy, a community-developed core Python package for Astronomy \citep{2018AJ....156..123A};
\end{itemize}
\label{lastpage}

\bibliographystyle{aa}
\bibliography{SDRref}
\begin{appendix}
\section{Post-processing filtering\label{sec:ppf}}
Although the pipeline described in Sect. 2 employs several quality assurance criteria, it is not able to remove completely the spurious periods occurring in the {\tt best\_rotation\_period} distribution (see discussion in Sec. \ref{sec:prot} and Fig. \ref{newdensity}). In order to understand the origin of these peaks in the  {\tt best\_rotation\_period} distribution, we performed a deep investigation of \gaia time-series and identified two new criteria to filter out the sources affected by the spurious signals. These criteria are based on the analysis of the per-transit corrected excess flux  time-series and on the image parameter determination (IPD) harmonic model time-series and will be discussed in the next subsections.
\subsection{The per-transit corrected excess flux}
The corrected excess flux $C^{*}$ is a parameter defined by \cite{2021A&A...649A...3R} in order to measure the consistency between the \g, the \bp and the \rp cumulated photometry for a given source.
\cite{2021A&A...649A...3R} defines firstly the excess flux factor $C$ as:
\begin{equation}
\label{excess factor}
C=\frac{I_{G_{\rm BP}} + I_{G_{\rm RP}}}{I_G}
\end{equation}
where $I_{G_{\rm BP}}$, $I_{G_{\rm RP}}$ and $I_{\rm G}$ are the cumulative fluxes in the \bp, \rp and \g bands, respectively.
Given the instrumental response and the  profiles of the \gaia passbands, $C$ should be slightly larger than the unity for a single and isolated main-sequence star.  In reality, $C$ has a colour dependency fitted by a polynomial function f(\bprp) whose coefficients are provided in the Table 2 of \cite{2021A&A...649A...3R}.
The corrected excess factor $C^{*}$ for a given source is then defined as the difference between the measured  and the expected C:
\begin{equation}
\label{correctedexcessfactor}
C^{*}=C-f(I_{G_{\rm BP}}- I_{G_{\rm RP}})
\end{equation}
A corrected excess factor significantly different from 0 may be an indicator of calibration issues (i.e. blending  or background over-estimation) or of a peculiar source (i.e. an extended source or a star with strong emission lines in the region where the \g and the \rp bands have different sensitivity).

We defined the per-transit corrected excess flux $c^{*}$ by applying  Eq. \ref{correctedexcessfactor} to the per-transit photometry.
In such a way, we obtained for each source a $c^{*}$ time-series that permits to investigate if the consistency between the three photometric fluxes exhibits some time dependence that could affect the data and explain the detection of the spurious periods.

\begin{figure}
\begin{center}
\includegraphics[width=80mm]{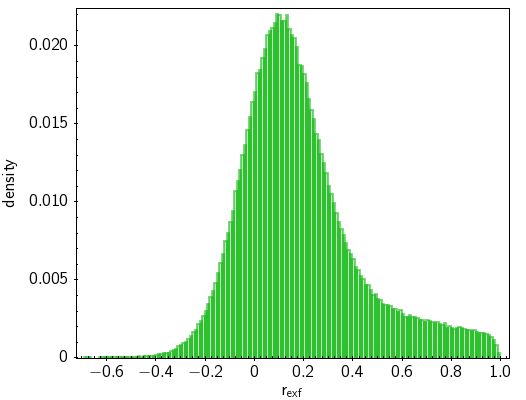}
\caption{Distribution histogram of the Spearman correlation coefficient $r(G,c^{*})$ between the $G$ magnitude and the corrected excess factor $c^{*}$. 	}
\label{spearmandistribution}
\end{center}
\end{figure}

\begin{figure}
\begin{center}
\includegraphics[width=80mm]{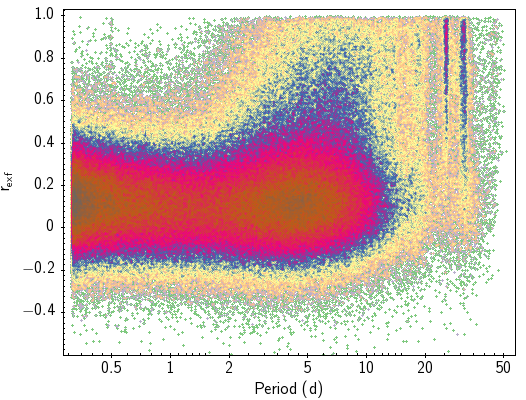}
\caption{Density distribution of the correlation coefficient $r_\text{exf} =r(G,c^{*})$ vs. the detected rotation period. The spurious periods around 25 and 32~d are characterized by an high correlation coefficient.} 
\label{perspear}
\end{center}
\end{figure}

In order to individuate the problematic sources, we   focussed our analysis on the correlation between  \g and $c^{*}$.
Indeed for isolated, single, main sequence sources no significant correlation is expected. 
For each source, we computed the Spearman correlation coefficient $r_\text{exf} =r(\g,c^{*})$ between the \g measurements and the $c^{*}$ values in order to test if the two quantities are connected by a monotonic relationship.
In Fig. \ref{spearmandistribution}, we display the distribution of $r_\text{exf} $ for all the stars detected by the pipeline. 
Although the peak is around 0, the distribution is moderately skewed with a tail towards positive values. 
We investigated the relationship between the detected periods and the $r_\text{exf} $ values and we found that
most of the sources with spurious periods are located in the tail of the $r_\text{exf} $ distribution.  This can be easily seen in Fig. \ref{perspear}, which reports the $period-r_\text{exf} $ density diagram. Indeed, the sources with 
spurious periods are mainly concentrated in the region of the diagram corresponding to $r_\text{exf}  >0.5$.

We visually inspected \g, \bp, \rp photometric time-series and \bp, \rp epoch spectra for tens of stars having $r_\text{exf} $ > 0.5 and we found that the high correlation index can be due to three different reasons:
\begin{enumerate}
\item{the star is a partially resolved binary;}
\item{the star has some calibration issue due to the transition from a window-gate configuration to another}
\item{the star has strong emission lines whose intensity is correlated with \g.}
\end{enumerate}

\begin{figure*}
\begin{center}
\includegraphics[width=160mm]{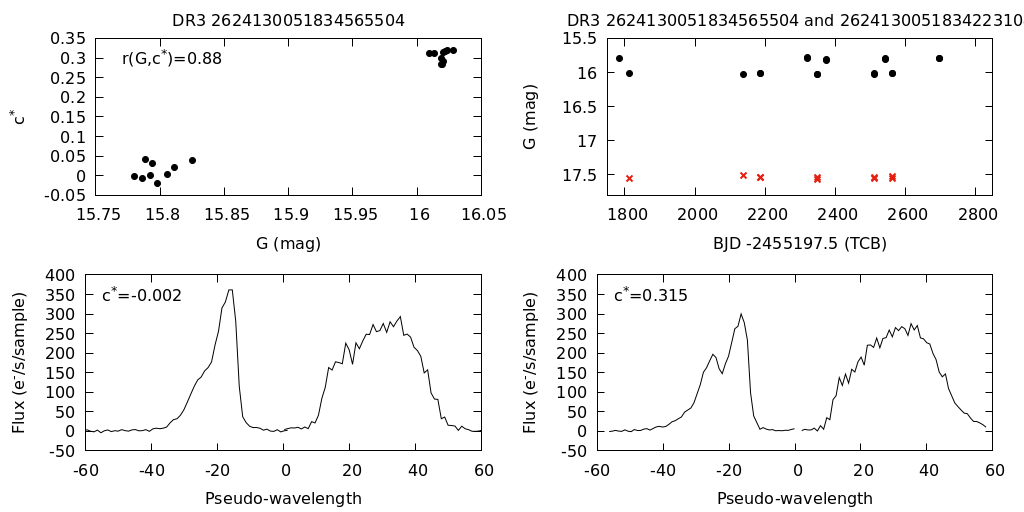}
\caption{Top left panel: $c^{*}$ vs. $G$ for the star {\tt DR3  2624130051834565504}: the data are distributed around two clumps, one centred around $c^{*}\simeq 0$ and the second around $c^{*}\simeq 0.315$. Top right panel: $G$ magnitude time-series of the star {\tt Gaia DR3  2624130051834565504} and of its close neighbour {\tt Gaia DR3 2624130051834223104}. The two sources form a partially resolved binary i.e. depending on the observation time and on the scan angle, the two stars can be resolved or unresolved in the astrometric instrument. Bottom left panel: \bp and \rp epoch spectra corresponding to the case in which the two stars are unresolved both in the astrometric and spectrophotometric instrument. In this case $c^{*}$ is close to 0 because both instruments collect the fluxes coming from the two stars.
Bottom right panel: \bp and \rp epoch spectra corresponding to the case in which the stars are resolved in the astrometric field but blended in the \bp and \rp FoV. In this case the source exhibits  a strong excess factor because the astrometric instrument collects the flux coming from a single source whereas the spectrophotometric instrument collects the flux coming from the star and its neighbour. Note that the \bp and \rp epoch spectra are here shown only for illustration. They will be available to the community in \gaia DR4. \label{partiallyresbin}  }
\end{center}

\end{figure*}

In Fig. \ref{partiallyresbin} we illustrate an example of the first kind of source. In the top left panel of the picture we display  $c^{*}$ vs. \g 
for the star {\tt Gaia DR3  2624130051834565504}. The two quantities are highly correlated with a Spearman coefficient $r(c^{*},\g) = 0.88$. 
The data are clustered in two different groups: at the brightest \g magnitudes the $c^{*}$ parameter is close to 0, whereas at the faintest magnitudes $c^{*}$ is around 0.3. 
The visual analysis of the epoch spectra time-series  reveals that the star is blended with another source that we identified as the star {\tt Gaia DR3 2624130051834223104} located at a distance of $0.6~arcsec$ from it.
When $c^{*}$ is close to 0, the two stars are unresolved by both the astrometric and the spectrophotometric instrument: in this case 
 the two instruments measure the total flux coming from the two stars and $I_{ G}$  is consistent with the sum of $I_{G_{\rm BP}}$ and $I_{G_{\rm RP}}$.  When $c^{*}$ is close to 0.3 the two stars are fully resolved by the astrometric instrument and partially resolved by the spectrophotometric instrument: in this second case the astrometric instrument provides  distinct measurements for  the two stars, whereas the spectrophotometric instrument attributes all the measured flux only to {\tt Gaia DR3 2624130051834565504} \footnote{In DR3, transits affected by blending are simply flagged. In DR4 a de-blending procedure will be put in place.} that, therefore, turns out to have an excess of $I_{G_{\rm BP}}$ and $I_{G_{\rm RP}}$ fluxes. In the bottom panels of Fig. \ref{partiallyresbin}, we display two epoch spectra corresponding to the two different $c^{*}$ values. When  $c^{*}$ is close to 0.3 the \bp spectrum exhibits two distinct peaks that correspond to the different sources: in this case the star is partially resolved in the \bp epoch spectrum and unresolved in the \rp epoch spectrum. When $c^{*}$ is close to 0, the star is unresolved in both epoch spectra. 
 In the top right panel we show the \g time-series of the two  stars: the black bullets  are used to mark the \g measurement of {\tt Gaia DR3  2624130051834565504} whereas the red crosses the \g measurements of {\tt Gaia DR3 2624130051834223104}. When the stars are unresolved \g gets brighter for {\tt Gaia DR3  2624130051834565504} and no measurements are supplied for  {\tt Gaia DR3 2624130051834223104}. 
In such a case, the alternation between measurements in which the sources are fully resolved and measurements in which they are unresolved depends on the scan angle under which the two sources are imaged \citep[see][for further details]{DR3-DPACP-162}

\begin{figure*}
\begin{center}
\includegraphics[width=160mm]{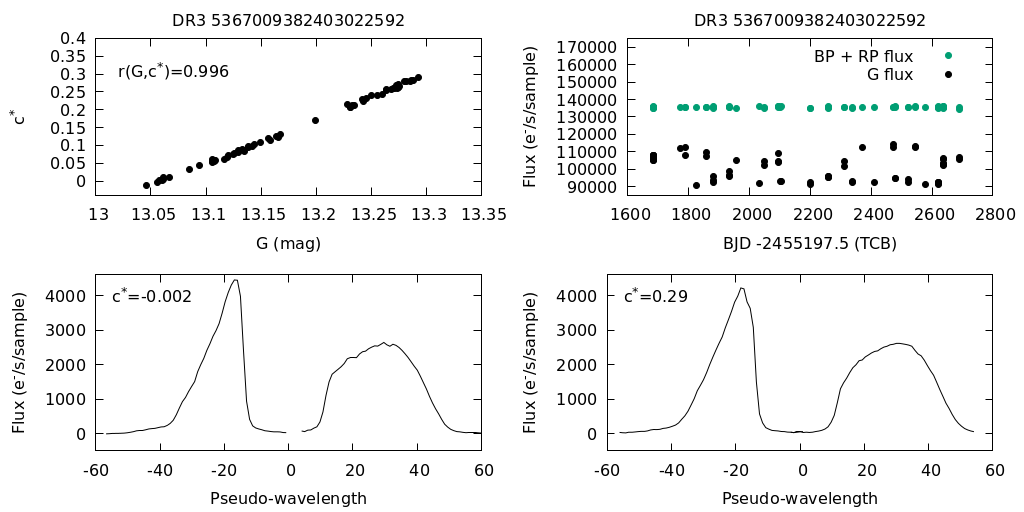}
\caption{Top left panel: $c^{*}$ vs. $G$ for the star {\tt Gaia DR3 5367009382403022592}. The two quantities are strongly correlated ($r_\text{exf} =0.996$). Top right panel: $I_G$  (black bullets) and $I_{G_{\rm BP}} + I_{G_{\rm RP}}$ (green bullets) time-series.	While $I_{G_{\rm BP}} +I_{G_{\rm RP}}$ is almost constant vs. time, the $I_G$ time-series exhibits temporal variations with a peak-to-peak amplitude of about 20\%. This variability is very likely due to instrumental issues otherwise it should be visible also in the $I_{G_{\rm BP}} + I_{G_{\rm RP}}$ time-series. Bottom left panel: \bp and \rp epoch spectra corresponding to $c^*=0.002$. Bottom right panel:\bp and \rp epoch spectra corresponding to $c^*=0.29$.  Note that the \bp and \rp epoch spectra are here shown only for illustration. They will be available to the community in \gaia DR4. }
\label{saturated}
\end{center}
\end{figure*}

In Fig. \ref{saturated} we illustrate an example of the second kind of source. In the top-left panel of the picture we plot \g vs. $c^{*}$ values for the star {\tt Gaia DR3 5367009382403022592}. The two quantities are highly correlated with a Spearman coefficient $r(c^{*},\g)~=0.99$. In the bottom panels we display the epoch spectra corresponding to $c^{*}~=0.002$ and $c^{*}=0.29$. In this star the spectra do not exhibit significant differences that could explain the high value of the Spearman coefficient. In the top-right panel of the picture we present instead the time-series of $I_{G}$ and of the sum $I_{G_{\rm BP}} +I_{G_{\rm RP}}$: the sum of \bp and \rp fluxes is almost constant whereas the $I_G$ flux exhibits time variations of the order of 10\%. In this source, the high value of the Spearman coefficient is due to the fact that, being the $I_{G_{\rm BP}} +I_{G_{\rm RP}}$ sum constant, $c^{*}$ is directly proportional to the $I_G$ flux. The \g variability is probably due to the fact that the star flux has been measured with different gate-windows configurations that, in some cases, could have different instrumental responses. Indeed, the star has a magnitude of about 13 mag that is the value at which the transition between the  2D and the 1D observation window occurs.
The signal detected in this star is, therefore, probably induced by the switching between the different configuration of the photometric instrument.

\begin{figure}
\begin{center}
\includegraphics[width=80mm]{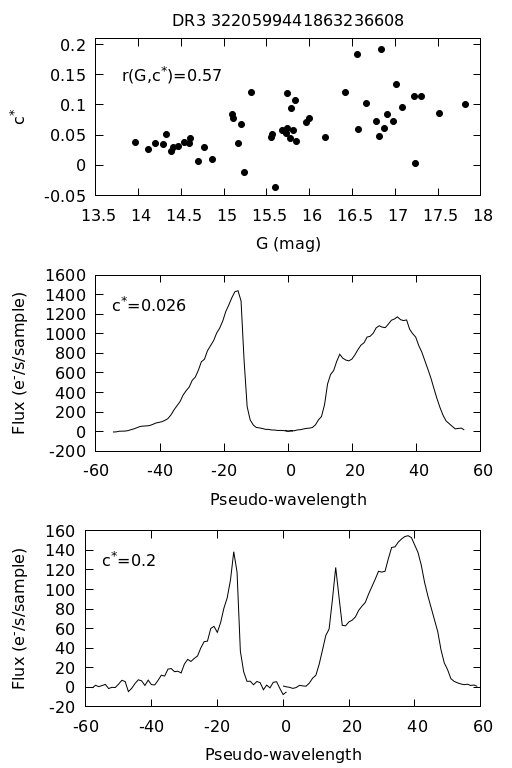}
\caption{Top panel: $c^{*}$ vs. $G$ for the star {\tt Gaia DR3 3220599441863236608}. The two quantities are moderately correlated ($r_\text{exf} =0.57$).
Middle panel: \bp and \rp epoch spectra corresponding to $c^{*} =0.026$. Bottom panel: \bp and \rp epoch spectra corresponding to $c^{*} = 0.2$. In this case the high corrected excess factor is due to a strong emission line falling in a region where the sensitivity of the $G$ and \rp bands are different. 
Note that the \bp and \rp epoch spectra are here shown only for illustration. They will be available to the community in \gaia DR4.
}
\label{emission2}
\end{center}
\end{figure}

\begin{figure}
\begin{center}
\includegraphics[width=80mm]{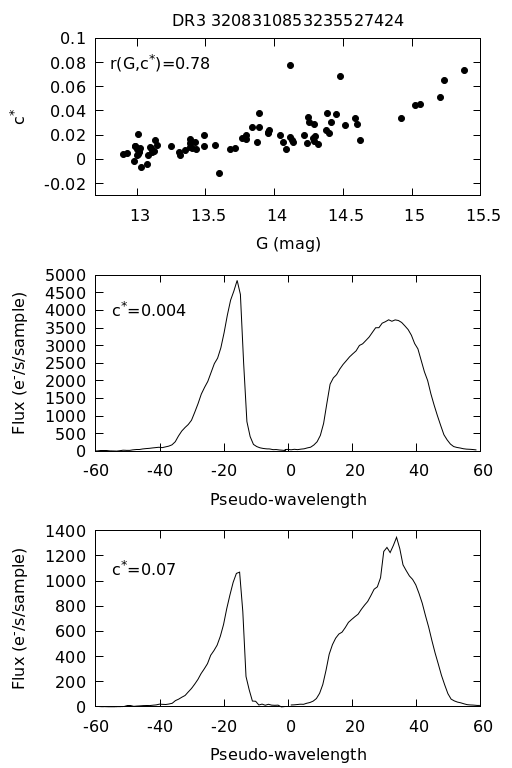}
\caption{Same of Fig. \ref{emission2} for the star {\tt Gaia DR3 3208310853235527424}.	}
\label{emission}
\end{center}
\end{figure}

Finally in Fig. \ref{emission2} and Fig. \ref{emission} we present two examples  of the third kind of star. In the top panel of Fig. \ref{emission2} we report $c^{*}$ vs. \g for the source {\tt Gaia DR3 3220599441863236608}. The two quantities are moderately correlated with a Spearman coefficient $r(G,c^{*})=0.57$. In the medium and bottom panels of the picture we display the epoch spectra corresponding to  two different transits having $c^{*}=0.026$ and $c^{*}=0.2$, respectively. In the transit with the higher $c^{*}$ value, the \rp epoch spectrum exhibits a strong emission line located in a region where the transmissivity of the \rp passband is higher than the \g one. This explains the excess of $I_{G_{\rm BP}}+ I_{G_{\rm RP}}$ with respect to $I_G$. A similar  situation is illustrated in Fig.  \ref{emission} for the star {\tt Gaia DR3 3208310853235527424}. In these two latter stars, the variability detected in the \g time-series is due to physical reasons and not to instrumental issues.

\subsection{Spurious periods induced by the scan angle }
\begin{figure}
\begin{center}
\includegraphics[width=80mm]{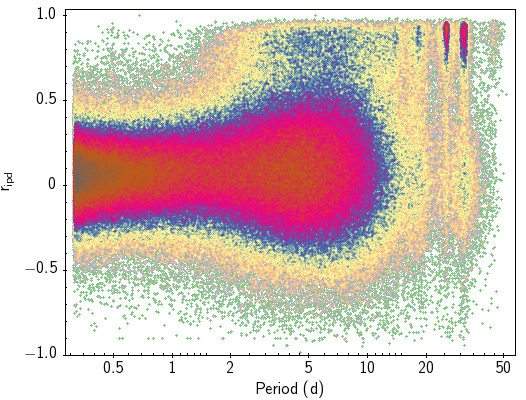}
\caption{Density map of the coefficient $r_\text{ipd}=r(hm(\psi), G)$ vs. the detected rotation period.	}
\label{r2periods}
\end{center}
\end{figure}

The procedure of PSF (Point Spread Function) and LSF (Line Spread Function) fitting of the \gaia pipeline is optimised for point-like sources \citep{2021A&A...649A...3R}. In the case of extended sources like galaxies or elongated sources like partially resolved binaries, the computation of  \g, \bp and \rp values can be affected by the scan direction $\psi$ of the satellite. For each \gaia source, the Goodness-Of-Fit (the reduced chi-square) of the IPD \citep[Image Parameters Determination][Sect.\ 3.3.6]{2021gdr3.reptE...3C} is fitted with an harmonic model $hm(\psi)$\citep[][Sect.\ 5]{2021A&A...649A...2L}:

\begin{equation}
\ln( GoF) =  hm(\psi)=c_0 + c_2\cos(2\psi) + s_2 \sin(2\psi)
\end{equation}

whose amplitude is given by
\begin{equation}
{\tt ipd\_gof\_harmonic\_amplitude }=\sqrt{(c_2^2 + s_2^2)}
\end{equation}
The higher the amplitude of the harmonic model, the higher the probability that the source is an extended source or partially resolved binary and that \g, \bp and \rp magnitudes are affected by $\psi$.
In order to quantify how much the \g band signal is affected by the variation in the IPD GoF  as a function of scan angle, Holl et al. 2022  computed the Spearman correlation coefficient between the \g band signal as a function of scan angle, and the IPD GoF model sampled at the observed scan angles:

\begin{equation}
r_\text{ipd}=r\left( \ { \{ \, G_i(\psi_i),\ hm(\psi_i) \ | \ i \in {1,...,N} \, } \ \}  \ \right)
\end{equation}
with $i$ being the observation index of a source having a total of $N$ observations, and $\psi_i$ being the associated scan angle.
In sources with a high $r_\text{ipd}$ value the \g time-series variation is coherently varying as a function of scan-angle and has a phase similar to the IPD GoF model, suggesting it is strongly affected by a scan angle dependent signal and thus could lead to the detection of spurious periods (further details on $r_\text{ipd}$ can be found in Holl et al. 2022.)
In Fig. \ref{r2periods} we show the density plot of $r_\text{ipd}$ vs. {\tt best\_rotation\_period} for the sourced detected by our pipeline. Also in this case, the spurious periods at 18, 25 and 32~d tend to be concentrated at higher $r_\text{ipd}$ values $(r_\text{ipd} >0.8)$.

\subsection{Filtering criterion}
\begin{figure}
\begin{center}
\includegraphics[width=80mm]{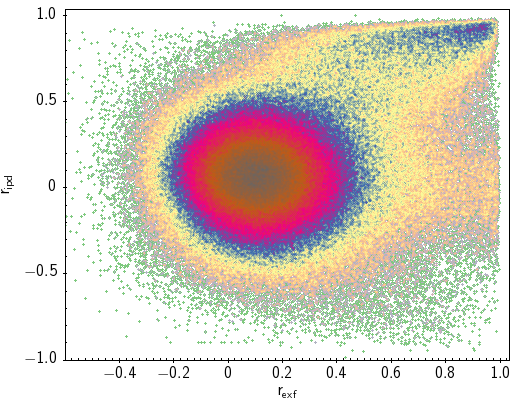}
\caption{Density map of $r_\text{ipd}$~vs.~ $r_\text{exf} $.	}
\label{r1r2}
\end{center}
\end{figure}

In order to avoid that the periods reported in our catalogue were affected by the issues illustrated in the above sections, we decided to discard all sources satisfying the condition:

\begin{equation}
\label{filter}
| r_\text{exf}  | >0.7   \mbox{~or~} |r_\text{ipd}| > 0.7 .
\end{equation}
We decided to use both indexes because they are complementary as illustrated in Fig. \ref{r1r2}.
 For a considerable number of stars the two indexes are correlated and high $r_\text{exf} $ values correspond to high $r_\text{ipd}$ values, but there is a subset of stars having $r_\text{ipd}$ values close to 0 and high $r_\text{exf} $ values.
This is the case, for instance, of the star {\tt Gaia DR3 5367009382403022592} reported in Fig. \ref{saturated}. This star, as discussed in the above section, has been observed in time with different window-gate configurations and suffers of calibration issues that give a $r_\text{exf} $ index close to 1.
The $r_\text{ipd}$ index is instead very low ($r_\text{ipd}=0.2$) because the source is an isolated single star and the PSF-fitting is not affected by the scan angle direction.
After applying the filter given by Eq. \ref{filter}, we obtained a final sample of 474\,026 bona-fide stars with rotational modulation.
The period distribution of these sources is displayed in Fig. \ref{finaldensity}. Most of the peaks associated with spurious periods are removed or considerably attenuated.

\section{Validation and quality assessment of the data }
\label{qualityassesment}
As mentioned in the above section, the {\tt gdr3\_rotmod} catalogue reports 474\,026
sources whose variability is likely induced by stellar magnetic activity. In order to estimate the percentage of the new discovered variables, we cross-matched the {\tt gdr3\_rotmod} with the \gaia DR3 catalogue of known variable stars \citep[hereafter {\tt gdr3\_xm}[]{DR3-DPACP-177}. This latter, compiled on the basis of  152 catalogs, reports the \gaia counterparts for a list of about 4.9 million of variable sources known from the literature. The cross-match between the {\tt gdr3\_rotmod} and the {\tt gdr3\_xm} catalogs turned out into 42306 variables.
Hence about 430000 sources reported in the {\tt gdr3\_rotmod} catalogue are new discovered variables.

In this section we assess the quality of the {\tt gdr3\_rotmod} catalogue by estimating the detection efficiency, the contamination and the percentage of correct rotation periods.  Finally a comparison between the {\tt gdr3\_rotmod} and the {\tt gdr2\_rotmod} catalogs is also carried out and discussed.
\subsection{Detection efficiency}
\label{sec::completeness}
In principle, all the stars with a  magnetic activity are characterized by a rotational modulation signal. An estimate of the survey completeness therefore could be given by the equation:
\begin{equation}
C=\frac{N_{\rm rot}}{{N_{\rm sel}\times \rm CF}}
\end{equation}
where $N_{\rm rot}$ is the number of stars for which the pipeline detected a rotational modulation signal, $N_{\rm sel}$ is the number of sources selected by the pipeline i.e. those falling in the HR region marked in  Fig. \ref{hr}, and $\rm {CF}$ is a correction factor that should take into account the following issues:
\begin{itemize}
\item {the selection sample can miss a certain percentage of good candidates because of the 20\% threshold in the relative error on parallax and because of the sharp edges of the selection region (see Fig. \ref{hr});  }
\item{the selection region can host contaminant sources;}
\item{the photometric \gaia completeness changes with the stellar magnitude and with the crowding of the scanned region. }
\end{itemize}
Given the above issues, making a precise estimate of CF is very difficult. Indeed it would require an "aprioristic" knowledge of how magnetically-active stars are distributed in our Galaxy and a detailed knowledge of the \gaia photometric completeness. This latter, is known only on a large scale \citep[see discussion in Sec. 6 of][]{2021A&A...649A...1G}.

For all  these reasons, in the present paper, we  limit ourselves to estimate only the detection efficiency of the pipeline that is given (in percentage unit) by:
\begin{equation}
E=100\times\frac{N_{\rm rot}}{N_{\rm sel}}  \simeq 0.4 \%.
\end{equation}

The detection efficiency is a complex function of the stellar magnitude, the amplitude of the rotational modulation, the stellar rotation period and the ecliptic latitude $\beta$. Indeed, as discussed in \cite{2012MNRAS.421.2774D},
the detection of the stellar rotation period is favored at certain ecliptic latitudes and, in general, biased towards shorter rotation periods ($P_{\rm rot} \le 5 d$).

We studied how $E$ depends on ecliptic latitude by binning the $N_{\rm sel}$ sources detected by the pipeline in $1~deg$-intervals.
For each interval, we computed the efficiency $e(\beta)$ as
\begin{equation}
 e(\beta)=100\times\frac{n_{\rm rot}(\beta)}{n_{\rm sel}(\beta)}   
\end{equation}
where $n_{\rm rot}(\beta)$ is the number of sources detected in the interval $(\beta-0.5\degree,\beta+0.5\degree)$ and $n_{\rm sel}(\beta)$ the number of sources selected in the same interval. 
In the same way we studied how E depends on the \g magnitude by binning the data in $0.5 mag$-intervals and computing  the efficiency $e(\g)$ as:
\begin{equation}
\rm{e}(G)=100 \times \frac{n_{\rm rot}(G)}{n_{\rm sel} (G)}
\end{equation}

\begin{figure}
\begin{center}
\includegraphics[width=80mm]{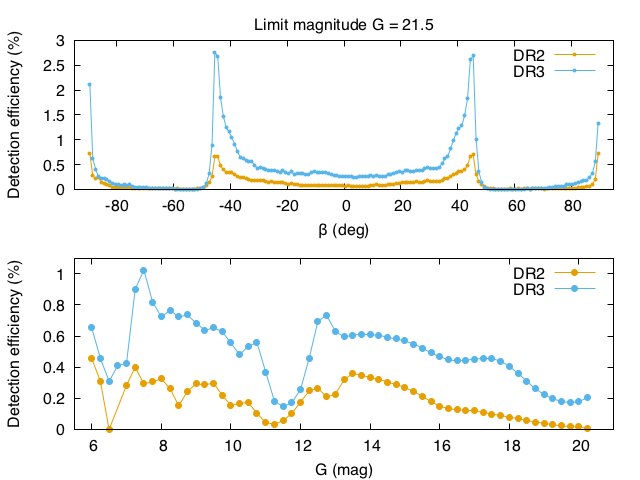}
\caption{Top panel. Detection efficiency vs. ecliptic latitude for DR3  (cyan line) and DR2 release (yellow line). 
Bottom panel. Detection efficiency vs. $G$ magnitude. }
\label{completeness}
\end{center}
\end{figure}

In the top panel of Fig. \ref{completeness} we plot  the trend of the detection efficiency vs. $\beta$ (cyan bullets): 
${e}(\beta)$ is almost constant ($\simeq 0.03\%$) in the latitude range $(-30\degree,30\degree)$,  increases up to 3\% at $\beta = \pm 45 \degree$ and then abruptly decreases and reaches the minimum value of 0.004 \% at $\beta=\pm 55 \degree$. Finally it increases again up to  2\% value at ecliptic poles.
This trend is  a direct consequence of the Gaia scanning law that is strongly dependent on ecliptic latitude and determines a better sampling around ecliptic poles and around the latitudes $\beta=\pm 45\deg$ (see Fig. \ref{mapavglen} and \ref{mapavgnp} of the present paper and the discussion in \cite{2012MNRAS.421.2774D} for further details).
The detection efficiency ${e}(\beta)$ of the {\tt gdr2\_rotmod} catalogue is also plotted (yellow bullets) for comparison.

\begin{figure}
\begin{center}
\includegraphics[width=80mm]{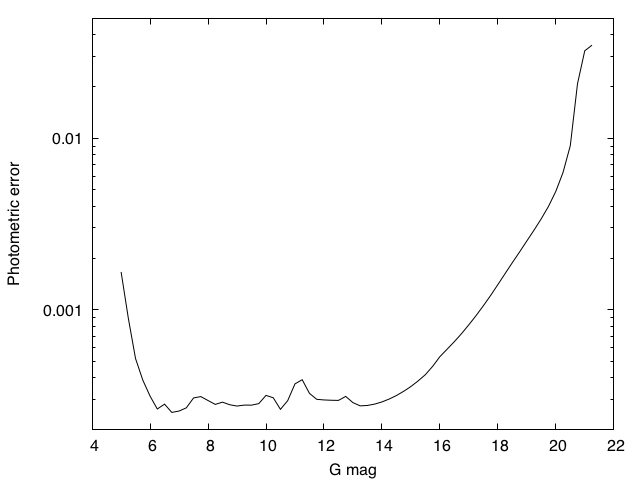}
\caption{Average photometric error vs. the stellar $G$ magnitude.	}
\label{err}
\end{center}
\end{figure}

In the bottom panel of Fig. \ref{completeness}, we plot how the detection efficiency changes with \g : in the \g range (13,20.5) the trend of $ {e}(G)$ is very regular and it gentle decreases with increasing magnitude. In the \g range (6,13) the dependency on \g is more complex and reflects the trend of the \gaia photometric uncertainties in that magnitude range.  In Fig.\ref{err} we display the average value of the \g photometric uncertainties for the sources selected by the pipeline. The uncertainty curve exhibits two peaks at $G\simeq 6.5~mag$ and at $G\simeq11.5~mag$: at the same magnitudes, $e(G)$ has two depth minima. The irregular trend of the \gaia photometric uncertainty in the (6,13) mag range is due to the switch between the different window-gate configurations adopted by \gaia to acquire the photometric measurements in different magnitude regimes \citep[see][for further details]{2021A&A...649A..11R,2021A&A...649A...3R}. In particular, the decreasing of the detection efficiency around \g = 11.5 mag, could be due to the mismatch between the rate at which charges are transferred in the along-scan direction and the rate at which 
the stellar image drifts across the CCD. This mismatch distorts the \gaia PSF  but, currently, this is not taken into account in the PSF modeling and determines an increase of the photometric uncertainty in that magnitude range \citep[][]{2021A&A...649A..11R}. In \gaia DR4 the PSF modeling will take this effect into account and a better photometry is expected.

Note that, while the detection efficiency obtained in DR3 is in general higher than in DR2, at \g $\simeq 11.5$ the two releases have about the same efficiency. This will be discussed in detail in Appendix \ref{dr2dr3}.

In the top panel of  Fig. \ref{completeness1315},
we show how the detection efficiency depends on $\beta$ down to the limiting magnitude \g=15.
In this case $e(\beta) \simeq 15\%$ at ecliptic poles and about 28 \% around $\beta = \pm 45 \deg $. If we restrict the analysis to the magnitude range (13:15) (where the average photometric  uncertainty is about 0.001 $mag$), we find that $e(\beta)$ is about 20 \% at ecliptic poles and about 33 \% around $\beta=\pm 45 \deg$.

\begin{figure}
\begin{center}
\includegraphics[width=80mm]{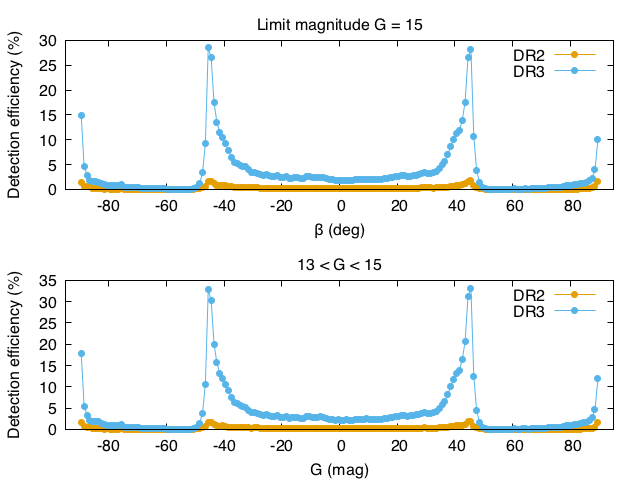}
\caption{Top panel: Detection efficiency vs. ecliptic latitude down to the limit magnitude $G = 15$.
Bottom panel: Detection efficiency vs. ecliptic latitude in the $G$ magnitude range (13:15).}
\label{completeness1315}
\end{center}
\end{figure}

\subsection{Contamination}
\label{contamination}
As discussed in \cite{2018A&A...616A..16L} for the {\tt gdr2\_rotmod} catalog, the possible contaminants of the {\tt gdr3\_rotmod}  could be given by pulsating variables for which parallaxes were not properly estimated or by short period eclipsing binaries. The sources currently listed in the    {\tt gdr3\_rotmod} do not overlap with any of the pulsating variables or eclipsing binaries reported in the \gaia DR3 \citep{DR3-DPACP-164}. Moreover all the quality criteria applied to filter out the data and described in Sections \ref{sec:prot} and \ref{sec:ppf} should remove pulsating variables whose light curves significantly deviate from the typical sinusoidal shape of rotational modulation variables \citep[see the discussion in][]{2018A&A...616A..16L}. So we expect that the contamination level of our catalogue is very low.

An accurate estimate of this contamination will require to know the variability classes for a subset of stars statistically representative of the full catalog. This is not possible because the currently available surveys of variable stars do not have the same properties of \gaia in terms of magnitude range and sky coverage. Moreover the variable classification performed by these surveys is in turn subject to some uncertainty.
Nevertheless, the comparison between the {\tt gdr3\_rotmod} catalogue and the available surveys can be useful to estimate at least the order of magnitude of the contamination level.

In order to have a rough estimate of the contamination level, we therefore cross-matched the {\tt gdr3\_rotmod}  with the ZTF (Zwicky Transient Facility)  catalogue of Periodic Variable Stars \citep[hereafter {\tt ztf\_var}][]{ztfcat} and with the  ASAS-SN (All Sky Automated Survey for SuperNovae)  catalogue of variable stars \citep [herafter {\tt asas-sn\_var}][]{asassncat}. 

\subsubsection{Comparison with ZTF classification}
The {\tt ztf\_var} has been obtained by analysing the photometric time-series acquired with the ZTF telescope \citep{ztftel} in the Sloan $g$ and $r$ photometric bands during a  470 d long survey. This survey \citep{ztfscience} covers the northern sky hemisphere and has a limiting magnitude $r \simeq 20.6$ that is similar to that of the {\tt gdr3\_rotmod}. The {\tt ztf\_var} catalogue lists 
781\,602 variable stars that have been classified in 11 different types. About 150\,000 of these stars have been  classified by the ZTF pipeline as rotational variables. 
The cross-match between the {\tt gdr3\_rotmod} and the {\tt ztf\_var} catalogue returned  6513   sources\footnote{The reader may wonder why a large fraction of the rotational variables reported in the {\tt ztf\_var} and in the {\tt asas-sn\_var} catalogs are missing in the {\tt gdr3\_rotmod}. This actually happens because most of these variables fall in sky regions poorly sampled by \gaia and do not satisfy the selection criteria to be processed by the pipeline (see Sect. \ref{datasel}).}   attributed by the \cite{{}} pipeline to 6 different classes. In  Table \ref{ztfclass} we reported the number and the percentage of the sources attributed to each class. Most part of the sources (94.6\%) were classified as rotational modulation variables according to the {\tt gdr3\_rotmod}. Indeed the 91.7\% of variables were classified as BY Dra and the 2.6\% as RS CVn variables. About 5 \% of the sources were classified as EW or EA binary systems.  A small percentage (0.4 \%) was attributed to the Semi-Regular class and finally only 2 stars out of the sample were classified as pulsating variables.

\subsubsection{Comparison with ASAS-SN classification}
The {\tt asas-sn\_var} catalogue has been obtained by analysing the $V-band$ photometric time-series collected by the ASAS-SN survey \citep{asassnsurvey} that covers the entire sky down to the limit magnitude $V~\simeq~17$. The time-series analysed to produce the catalogue were acquired between 2013 and 2018 and are characterized by a 2-d or a 3-d cadence. The catalogue reports 687\,695 sources and supplies a classification for 662\,627 of them (the remaining are classified as uncertain variables). About 90\,000 sources of the catalogue are classified as rotational variables. 

The sources common to the {\tt gdr3\_rotmod} and the {\tt asas-sn\_var} catalogue are 1\,054.
In Table \ref{asassnclass} we report the classes attributed by \cite{asassncat} to the 1054 stars$^7$ and their relative percentages.
The 80.25\% of the 1\,054 stars were classified as rotational variables. The 13 \% of the sources were classified as binary systems (most part of which are of EA type). Six sources were classified as Young Stellar Objects (YSO), 7 sources as pulsators, 20 sources as UV Cet stars and finally 22 sources, corresponding to the tag VAR, have an uncertain classification. Note that the UV Cet variables are magnetically active stars whose light curves are mainly characterized by outbursts due to energetic flare events. However these stars can also exhibit a rotational modulation signal, so they cannot be regarded as contaminants. In the same way the stars with uncertain classification could be magnetically active stars for which the ASAS-SN pipeline was not able to detect the rotational signature. Note that for these uncertain stars, the {\tt asas-sn\_var} catalogue usually reports period longer than 100 d that could be ascribed to short-term magnetic activity cycles like those detected by \cite{distefano2017}. In these cases, the evolution of MARs could have masked the rotational modulation signal and the asas-sn pipeline could have failed in the  detection of the rotation period because it does not adopt a segmentation strategy.    

Overall, the comparison with the {\tt ztf\_var} and {\tt asassn\_var} catalogs assesses the contamination level of the  {\tt gdr3\_rotmod} between the 6 and 14 \%. The main contaminants of the catalogue are given by binary systems, whereas  pulsators can be considered negligible.

\begin{table}

\caption{Variability classes attributed by \cite{ztfcat}  to the 6\,513 stars common to the {\tt ztf\_var} and {\tt dr3\_rotmod} catalog.   }
\centering
\begin{tabular}{lll}
\hline
Class  & \# sources &  \%   \\
\hline
 BY Dra  & 5\,978 & 91.7 \% \\           
 RS CVn& 170 &2.6\%\\   
 EW &307 & 4.7\%\\
 EA &32 & 0.5\%\\
 SR & 28& 0.4\%\\
 DSCT&1&0.01\%\\
  RR&1&0.01\%\\
\hline

\hline
\end{tabular}
\label{ztfclass}
\end{table}
\begin{table}

\caption{Variability classes attributed by \cite{asassncat}  to the 1\,054 stars common to the {\tt asas-sn\_var} and {\tt dr3\_rotmod} catalog.  }
\centering
\begin{tabular}{lll}
\hline
Class  & \# sources &  \%   \\
\hline
 ROT  & 851 & 80.7 \%   \\           
 EA &118 & 11.2\%  \\
 EW &10 & 0.9\%   \\
 EB & 10& 0.9\%  \\
 RRAB&2&0.2\%  \\
 RRC&2&0.2\% \\
VAR&34&3.2\%\\
YSO&7& 0.66\% \\
UV Cet & 20& 1.8\%\\
\hline
\hline
\end{tabular}
\label{asassnclass}
\end{table}

\subsection{Rate of correct periods detection}
\label{correctdetections}
The filtering procedures described in Secs. \ref{sec:ppf} and \ref{sec:prot}  leads us to reject about  40\% of the sources initially detected by the pipeline. We are therefore confident that most of the sources with spurious periods are removed from the final catalog. Nevertheless, a small percentage of aliases and artificial periods introduced by the peculiar Gaia sampling or by photometric artefacts can  still be present in the released data.
An estimate of the rate of correct period detection cannot be easily done because catalogs from other surveys are in turn affected by sampling or calibration issues. However, a comparison with other surveys can still be useful to assess a lower limit for such a rate.
\subsubsection{Comparison with ZTF periods}
\begin{figure}
\begin{center}
\includegraphics[width=80mm]{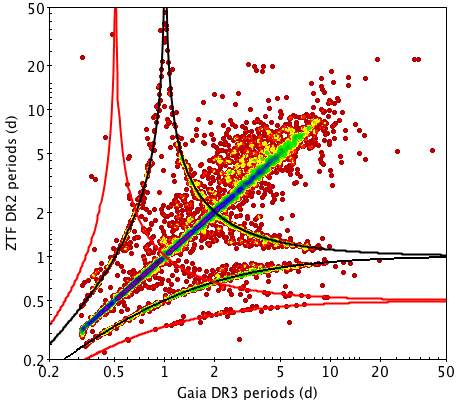}
\caption{Comparison between the periods detected in the ZTF survey and those detected in the Gaia DR3 release. The points are color-coded according to their density. The red and dark lines mark the loci defined by Eq. \ref{beatztf} for n=1 and n=2, respectively. \label{ztf}}
\end{center}
\end{figure}

In Fig. \ref{ztf}, we show a comparison between the  {\tt gdr3\_rotmod} and the {\tt ztf\_var} periods  for all the variables common to the two surveys and classified as BY Dra or RS CVn in the ZTF catalog. 
Two periods can be considered in agreement if they differ  by less than 20\%. This tolerance value has been set from the work of \cite{distefano2017}. In fact, that paper demonstrates that the stellar rotation period detected for a given star can noticeably change from an observational season to another. They attributed these variation to the combined effect of Scwabe-like cycles and Surface Differential Rotation. However, \cite{2020ApJ...901...14B} pointed out that these variations are more likely due to the appearing and disappearing of MARs at random longitudes.
We found that the  Gaia and ZTF periods are in agreement in 68\% of the sample.

In  9\% of the cases, the Gaia periods are in agreement with half or twice the periods detected by ZTF. These mismatches are very common when dealing with rotational variables because, according to the longitudinal distribution of MARs, the light curves of these stars can exhibit one or two minima per rotation cycle. The long-term monitoring of solar-like variables performed by the {\it Kepler} mission showed that the light curve of a given star is often characterized by a continuous switch between a single-dip and a double-dip configuration (see \cite{2018ApJ...863..190B} and \cite{2020ApJ...901...14B} for details). Hence these mismatches reflect an  intrinsic properties of the rotational variables, and affect all the rotation period distribution collected by different surveys.

In 14\% of the sample, 
the ZTF periods and Gaia periods  are connected by a relationship of the type: 
\begin{equation}
\label{beatztf}
\frac{1}{P_{\rm ztf}}=\left| \pm \frac{1}{P_{\rm gaia}} \pm \frac{1}{ nT_{\rm ztf}}\right |
\end{equation}
where $T_{\rm ztf} = 1 d$ is the typical cadence of the ZTF survey and $n=1,2$. 
In these cases the periods detected by ZTF are aliases i.e. spurious periods generated by the interference between the sampling frequency and the rotational frequency (correctly detected by Gaia).
Note that, in principle, \gaia periods can also be affected by aliasing but \cite{2005MNRAS.361.1136E} showed that the irregularity of the \gaia scanning law  limits the building up of coherent signatures at regularly spaced frequencies. 

Finally in 9\% of the sample there is no evident relationship between the periods detected by Gaia and those detected by ZTF. 
\subsubsection{Comparison with asas-sn periods.}
\begin{figure}
\begin{center}
\includegraphics[width=80mm]{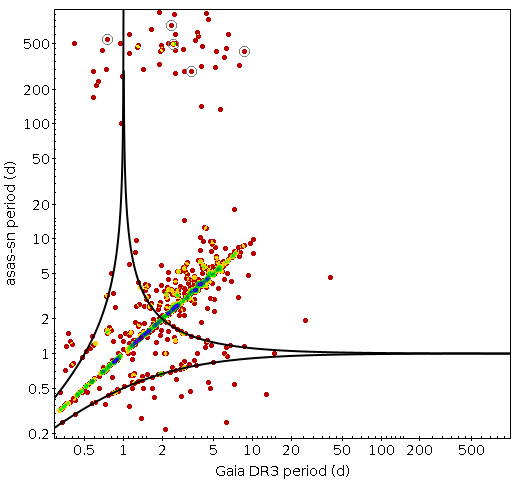}
\caption{Comparison between the periods detected in the asas-sn survey and those detected in the Gaia DR3 release. The points are color-coded according to their density. The continuous dark line depicts the loci defined by Eq. \ref{beatasas} for n=1}. The sources surrounded with the dark circles are also reported in \cite{hartman} with periods very similar to those detected in the Gaia DR3 (see Table \ref{asashat}).	\label{asassn}

\end{center}
\end{figure}

In Fig. \ref{asassn} we show the comparison between  the {\tt asas-sn\_var} and the {\tt gdr3\_rotmod} periods for all the stars common to the two surveys and labelled as rotational variables in the {\tt asas-sn\_var} catalog.

The \gaia periods agree with the asas-sn periods in 59\% of the sample and agree with half or twice the asas-sn periods in 12\% of the cases. In about 8\% of the cases the asas-sn periods fall in the locus defined by
\begin{equation}
\label{beatasas}
\frac{1}{P_{\rm asas}}=\left|\pm \frac{1}{P_{\rm gaia}} \pm \frac{1}{nT_{\rm asas}}\right|
\end{equation}
with n=1 and $T_{\rm asas} = 1 d$.
Finally in about 20 \% of the sample the \gaia periods do not match at all with the asas-sn periods.  

In conclusion, the comparison between \gaia and the two surveys permits to assess the rate of correct periods detection between 80 and 91\%. 

Actually, these percentages are lower limits for the rate of correct detection because the asas-sn and the ZTF pipelines do not take into account the intrinsic evolution of active regions and do not apply any segmentation strategy. Hence the two surveys, in some cases, are not able to pick the rotational modulation signal. This is the case, for instance, of the five stars encircled in Fig. \ref{asassn} for which instead \cite{hartman} reported rotation periods in good agreement with those found in the present work.

In table \ref{asashat} we reported the periods listed in the {\tt gdr3\_rotmod} and in the {\tt asas-sn\_var} catalogs and those found  \cite{hartman} for the five stars. The longer periods of the order of 200-500 d reported in the {\tt asas-sn\_var} are  probably due to long-term variations  of the stellar magnetic activity that in some cases can produce quasi-periodic signals \citep[see e.g.][and reference therein]{2016A&A...590A.133O,distefano2017}.

\begin{table*}
\label{asashat}
\caption{Comparison between {\tt gdr3\_rotmod}, {\tt asas-sn\_var} and {\tt HATNet} periods for the five stars encircled in Fig. \ref{asassn}.}
\begin{tabular}{|l|l|r|r|r|}
\hline
  \multicolumn{1}{|c|}{DR3 Id} &
  \multicolumn{1}{c|}{HATNet Id} &
  \multicolumn{1}{c|}{$P_{\rm asas}$ (d)} &
  \multicolumn{1}{c|}{$P_{\rm HATNet}$(d)} & 
  \multicolumn{1}{c|}{$P_{\rm gaia}$ (d)}\\
\hline
   417488275919654656 & HAT-087-000\,817\,9 & 426.13& 8.88 & 8.68\\
   391326775424158848 & HAT-087-002\,459\,6 & 544.9 & 0.75 & 0.75\\
   436212958941004032 & HAT-091-001\,087\,9 & 494.5 & 2.47 & 2.45\\
   219651916081208320 & HAT-169-000\,736\,4 & 285.15 & 3.41 & 3.38\\
   1927973852394016384 & HAT-205-002\,401\,2 & 716.4 & 2.34 & 2.34\\

\hline\end{tabular}
\end{table*}

\section{Comparison between DR3 and DR2 results}
\label{dr2dr3}
The detection efficiency of DR3 is about three times that of DR2 (see Fig. \ref{completeness}). However many of the rotational modulation variables reported in the {\tt gdr2\_rotmod} catalogue are not listed in the {\tt gdr3\_rotmod}.   
The {\tt gdr2\_rotmod} catalogue reports 147\,774 rotational modulation variables. The sources in common between the two catalogs are only 34\,747.  Hence 113\,027 stars listed in DR2 as rotational modulation variables do not appear in the DR3 catalog.

\begin{figure}
\begin{center}
\includegraphics[width=80mm]{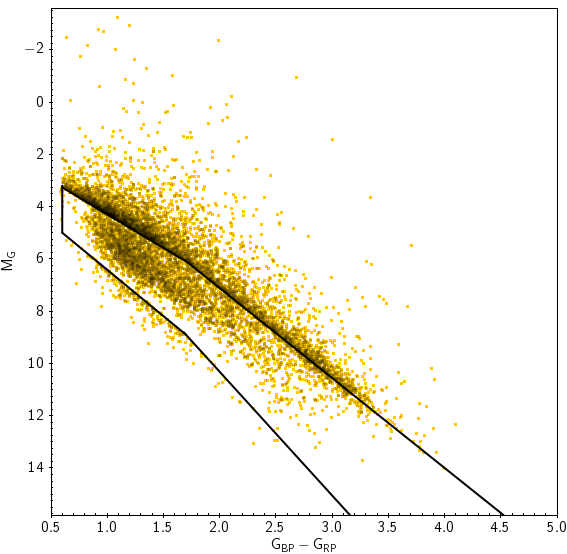}
\caption{Location of the DR2 variables, excluded by the DR3 selection criteria, in the $M_G~vs.~(G_{\rm BP}-G{\rm RP})$ diagram.}
\label{noselected}
\end{center}
\end{figure}

\begin{figure}
\begin{center}
\includegraphics[width=80mm]{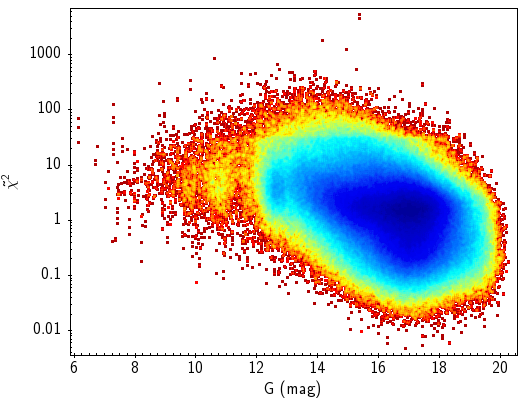}
\caption{Distribution of the $\tilde\chi^2$ values obtained by fitting the $G$ photometric time-series to the model given by Eq. \ref{model}.}
\label{chiredmag}
\end{center}
\end{figure}

\begin{figure}
\begin{center}
\includegraphics[width=80mm]{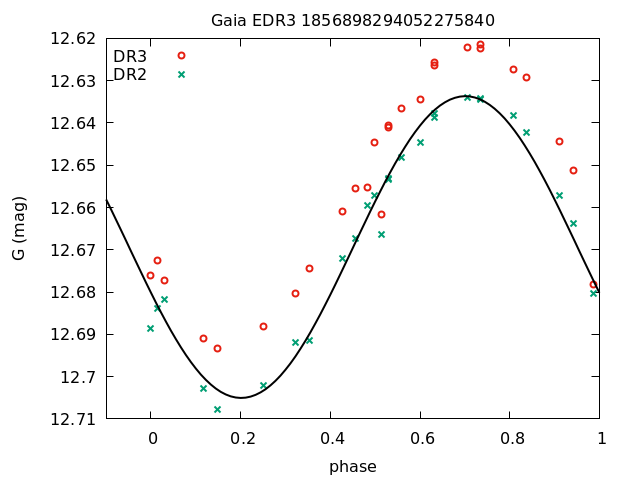}
\caption{Comparison between the DR3 (red circles) and DR2 (green crosses) photometric data  for the source {\tt Gaia DR3 1856898294052275840}. The two sets of photometric data correspond to the same set of transits but have different values because of the different calibration strategies applied in DR3 and in DR2. In this case the DR3 data have just a small offset respect to the DR2 data and the shape of the rotational modulation signal is preserved by the DR3 calibration.   }
\label{goodcal}
\end{center}
\end{figure}

\begin{figure}
\begin{center}
\includegraphics[width=80mm]{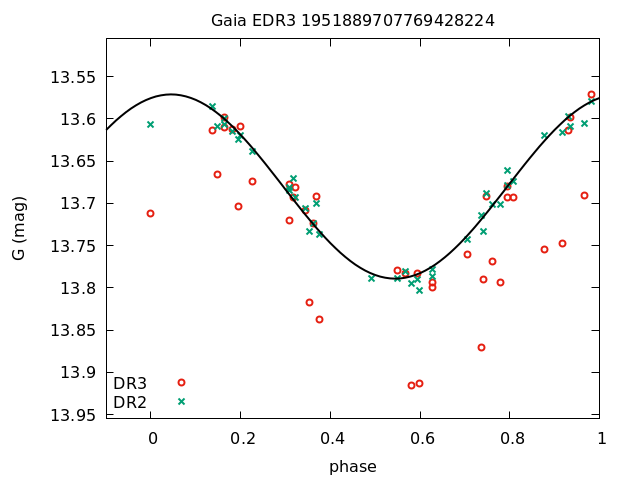}
\caption{The same of Fig. \ref{goodcal} for the source {\it Gaia DR3 1951889707769428224}. In this case the DR3 calibration distorces the rotational modulation signal detected in the DR2 photometry.}
\label{badcal}
\end{center}
\end{figure}

The missing stars have not to be considered as wrong DR2 detections. Indeed the contamination level and the rate of correct period detections, in this sample of stars, are of the same order of those seen in the  {\tt gdr3\_rotmod} catalogue (see Appendix \ref{qualityexluded} for details). 
These stars do not figure in the {\tt gdr3\_rotmod} catalogue only for technical issues that are here reported and discussed in details:
\begin{itemize}
	
    \item {{\tt selection issues}: the 8\% of the missing stars have different parallaxes and or mean magnitudes from those reported in DR2 and do not satisfy anymore one of the first two criteria reported in Sec. \ref{datasel}. In Fig. \ref{noselected}, we reported the location of these rejected  stars. A part a few hundreds of stars that now lie significantly distant from the main sequence region, the most part of them  (about 90 \%) still fall in the HR region used for selection  and have been rejected only because the relative error in parallax is greater than 20 \%;  }
    \item{{\tt segmentation issues}: as we stated in Sec. \ref{datasel}, the photometric time-series processed by the  CU7 pipeline are obtained by applying a chain of several operators to the \g raw time-series. This chain of operators, described in detail in \cite{DR3-documentation}, is slightly different from that employed in DR2 and adopts a different strategy to identify and remove possible outliers. In addiction, the pipeline used to detect rotational modulation adopts two further filters described  in Sec. \ref{sec:cleaning} that in DR2 were deactivated. 
    The combined action of CU7 operators and of the pipeline filters reduces
    the average number of points per segment and in turn  the number of segments available for the analysis. This issue affects the 38\% of the missing sources.
   These sources were discarded for two different  reasons:
   \begin{itemize}
       \item {the period search was not performed because the time-series segments (after filtering) have not enough points to be processed (we recall that the minimum number of points required to perform the period search is 12)}
       \item {the Gaia DR2 rotation period was correctly detected but only in one segment (we recall that a star is released if the same period is detected in at least two segments.} ) 
   \end{itemize}
   We  emphasize that the $G$ measurements removed by the cleaning procedure described in Sec. \ref{sec:cleaning}  are not necessarily bad points. For instance, the pipeline removes all
   the $G$ measurements for which the \bp or \rp counterparts are missing. This is done in order to assure that the activity indexes computed in the different bands are inferred from the same set of transits but, unfortunately turns out in the rejection of $G$ measurements suitable for the period search. 
   The pipeline removes also $G$ measurements corresponding to transits in which the \bprp colour significantly deviates from  the average \bprp and flags them as candidate flares. 
   A posteriori analysis of the data showed that  in the most part of these transits the anomalous colour is due to a bad measurement in the \bp or \rp band and does not correspond to an enhancement of the $G$ flux.
   In DR4 the pipeline will be refined to improve the cleaning strategy and minimise the loss of 
   valid measurements;

   }
  
    \item{{\tt quality criteria issues}:  37\% of the missing variables were correctly detected by the pipeline but were discarded because did not match one of the quality criteria described in Sec. 2.3. These sources can be divided  into three groups:
     \subitem {in a group of sources (13\% of the total sample) the period is correctly detected in all segments but the phase coverage criteria are not satisfied in any of the segments.  }
       \subitem {in a group of sources (14\% of the total sample) the period is correctly detected in all segments but the FAP exceeds the fixed threshold value.}
       \subitem{in a group of sources (10\% of the total sample) the period is correctly detected in all the segments but the reduced chi-square $\tilde\chi^2$ associated with the model-fitting is above the fixed threshold.}
       
       Also in this case, we emphasize that these discarded sources are not necessarily bad. Indeed the cleaning procedure described in the above issue reduces the number of points per segment and makes  the match of the phase coverage criteria more difficult than in DR2. 
       
       The reduced number of points per segments has also the effect to enhance the FAP value associated to a given period. Indeed the FAP formula employed  here \citep{2008MNRAS.385.1279B} is  dependent on the number  of points used to perform the period search.
       
       Finally, the analysis of the data suggests that the sources with an high $\tilde\chi^2$ value are probably stars for which the photometric errors have been under-estimated. The last problem mainly affects sources brighter than 13 mag. In Fig. \ref{chiredmag} we present the $\tilde\chi^2$ values computed by the pipeline according to  Eq. \ref{chi} versus the stellar magnitude. The picture shows that the $\tilde\chi^2$ values tend to decrease with increasing magnitudes. This trend suggests that photometric errors could be under-estimated for the brightest sources and over-estimated for the faintest sources. Unfortunately, this quality criterion caused the rejection of well known bright stars such as $AB~Dor$  and $BD-07 2388$;
       
       }
    
    \item{{\tt unknown issues}: for the 8\% of the missing sources we couldn't find a satisfying explanation. The  analysis of the data suggests that in some of these stars the rotational modulation signal could have been distorted by the strategy of the photometric calibration employed in DR3. 
    In Fig. \ref{goodcal} and \ref{badcal}  we show the comparison between the DR3 and DR2 $G$ time-series for two different stars. The two sets of measurements correspond to the same set of transits but assume different values because of the different calibration strategies. In both pictures, the two time-series are folded according to the rotation periods detected in DR2. In the first star, the DR3 photometric calibration induced a 0.02 mag offset between the two time-series but preserved the shape of the rotational modulation signal. In the second case, the shape of the rotational signal is well defined in DR2 time-series and strongly distorted  in the DR3 one.  (Note that the  periods detected in the DR2 time-series for the two stars are highly reliable because  they were detected also by \cite{asassncat} in the asas-sn time-series.)  }
\end{itemize}


\label{qualityexluded}
\subsection{Quality assessment of the DR2 excluded sources}
\begin{figure}
\begin{center}
\includegraphics[width=80mm]{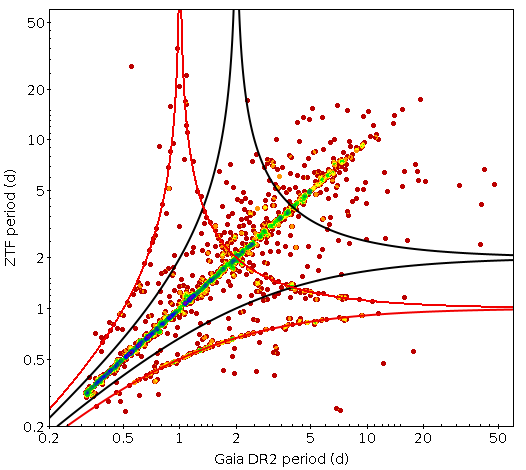}
\caption{Comparison between the periods detected in the ZTF survey and those detected in the \gaia DR2 release and excluded in the \gaia DR3 release. The points are color-coded according to their density.\label{dr2ztf}	}

\end{center}
\end{figure}

\begin{figure}
\begin{center}
\includegraphics[width=80mm]{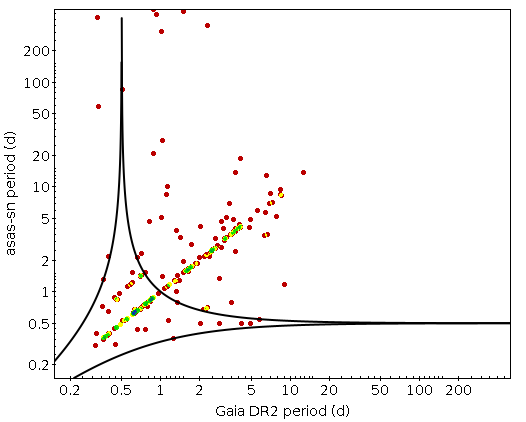}
\caption{Comparison between the periods detected in the asas-sn survey and those detected in the \gaia DR2 release and excluded in the \gaia DR3 release. The points are color-coded according to their density.	
\label{dr2asas}}
\end{center}
\end{figure}

We assessed the quality of the {\tt gdr2\_rotmod} sources that do not  appear  in the {\tt gdr3\_rotmod} catalog. Hereafter this sample of stars will be referred to as the {\tt gdr2\_excluded} list.
We evaluated the contamination level and the rate of correct period detection of this list by employing the same procedure adopted in Sects. \ref{contamination} and \ref{correctdetections} for the DR3 variables.
We cross-matched the {\tt gdr2\_excluded} list with the {\tt ztf\_var} and the {\tt asas-sn\_var} catalogs. We found 1\,692 sources in common with the {\tt ztf\_var} catalogue and 335 with the {\tt asas-sn\_var}. We reported the classes attributed to these stars by the ZTF and the asas-sn classifiers in Tables \ref{ztfclassdr2} and \ref{asassnclassdr2}, respectively. For each class, we reported the number and the percentage of the stars assigned to it. The  two Tables show that  contamination level of the {\tt gdr2\_excluded} can be assessed between 15 and 17\% and that, as in the case of the {\tt gdr3\_rotmod} catalog, the main sources of contamination is given by binary systems. 
We compared the periods of the {\tt gdr2\_excluded} list with those reported in the {\tt ztf\_var} and in the {\tt asas-sn} catalogs in order to evaluate the rate of correct period detections.
In Fig. \ref{dr2ztf} and \ref{dr2asas} we reported the comparison of the {\tt gdr2\_excluded} periods with the ZTF and the asas-sn periods, respectively.  In the sample of stars in common with ZTF, the 65\% of the {\tt gdr2\_excluded} periods differ by less than 20\% from the ZTF periods, the 12\% of the periods are compatible with half or twice the ZTF periods and the 17\% fall in the loci of aliases defined by Eqs. \ref{beatztf}.
In the sample of stars in common with asas-sn,  58\% of the {\tt gdr2\_excluded} periods differ by less than 20\% from the asas-sn periods, the 13\% of the periods are compatible with half or twice the asas-sn periods and the 4\% fall in the loci of aliases defined by Eqs. \ref{beatasas}.
\begin{table}

\caption{Variability classes attributed by \cite{ztfcat}  to the 1\,692 stars common to the {\tt ztf\_var} and {\tt gdr2\_excluded}  list.   }
\centering
\begin{tabular}{lll}
\hline
Class  & \# sources &  \%   \\
\hline
 BY Dra  & 1\,410 & 83.3 \% \\           
 RS CVn& 26 &1.5\%\\   
 EW &229 & 13.5\%\\
 EA &12 & 0.7\%\\
 SR & 10& 0.6\%\\
 DSCT&4&0.2\%\\
  RR&1&0.06\%\\
\hline

\hline
\end{tabular}
\label{ztfclassdr2}

\end{table}
\begin{table}

\caption{Variability classes attributed by \cite{asassncat}  to the 335 stars common to the {\tt asas-sn\_var} and the {\tt dr2\_rotmod} excluded list.  }
\centering
\begin{tabular}{lll}
\hline
Class  & \# sources &  \%   \\
\hline
 ROT  & 256 & 75.7 \%   \\           
 EA &13 & 3.85\%  \\
 EW &49 & 14.5\%   \\
 EB & 4& 1.2\%  \\
 RRAB&0&0\%  \\
 RRC&0&0\% \\
VAR&9&2.6\%\\
YSO&0& 0\% \\
UV Cet& 4& 1.2\%\\
\hline
\hline
\end{tabular}
\label{asassnclassdr2}

\end{table}
\end{appendix}

\end{document}